\documentclass[12pt,preprint]{aastex} 
\usepackage{float}
\usepackage{ulem}
 
\shorttitle{Pulsating EB KIC 9851944} 
\shortauthors{Guo et al.} 
 
\begin{document} 
 
\received{} 
\accepted{} 
 
\title{Kepler Eclipsing Binaries with $\delta$ Scuti/$\gamma$ Doradus Pulsating Components \uppercase\expandafter{\romannumeral1}: 
KIC 9851944 }  
 
\author{Zhao Guo, Douglas R. Gies, Rachel A. Matson} 
\affil{Center for High Angular Resolution Astronomy and  
 Department of Physics and Astronomy,\\ 
 Georgia State University, P. O. Box 5060, Atlanta, GA 30302-5060, USA; \\guo@chara.gsu.edu, rmatson@chara.gsu.edu, gies@chara.gsu.edu} 
 
\author{Antonio Garc\'{i}a Hern\'{a}ndez} 
\affil{Instituto de Astrof{\'i}sica e Ci\^{e}ncias do Espaco, Universidade do Porto, CAUP, Rua das Estrelas, PT4150-762 Porto, Portugal; \\
agh@astro.up.pt}

\slugcomment{accepted version} 

 
\begin{abstract} 
KIC 9851944 is a short period ($P=2.16$ days) eclipsing binary  in the {\it Kepler} field of view. By combining the analysis of {\it Kepler} photometry and phase resolved spectra from Kitt Peak National Observatory and Lowell Observatory, we determine the atmospheric and physical parameters of both stars. The two components have very different radii ($2.27R_{\odot}$, $3.19R_{\odot}$) but close masses ($1.76 M_{\odot}$, $1.79M_{\odot}$) and effective temperatures ($7026$K, $6902$K), indicating different evolutionary stages. The hotter primary is still on the main sequence (MS), while the cooler and larger secondary star has evolved to post-MS, burning hydrogen in a shell. A comparison with coeval evolutionary models shows that it requires solar metallicity and a higher mass ratio to fit the radii and temperatures of both stars simultaneously. Both components show $\delta$ Scuti type pulsations which we interpret as p-modes and p and g mixed modes. After a close examination of the evolution of $\delta$ Scuti pulsational frequencies, 
we make a comparison of the observed frequencies with those calculated from MESA/GYRE.

\end{abstract}


 
 
\section{Introduction}                              
The analysis of eclipsing binaries (EBs) with the simple geometric effect and Kepler's 3rd law offers us a means to determine accurate stellar masses and radii. Asteroseismology, the study of stars through their oscillation frequencies, also provides us with accurate stellar parameters. The pulsating eclipsing binaries are thus the best laboratories to test and refine our knowledge of stellar structure and evolution.

$\delta$ Scuti stars are radial and non-radial p-mode pulsators with frequency from $4$ to $60$ d$^{-1}$. They are within the mass range of $1.5$ to $2.5$ $M_{\odot}$ and of spectral type A2-F5. $\gamma$ Doradus variables are main sequence high order g-mode pulsators with pulsation period from $0.3$ to $3$ days. There are $\delta$ Scuti/$\gamma$ Dor hybrids, where both p and g-modes are present (Grigahc\`{e}ne et al. 2010; Uytterhoeven et al. 2011). Recent space observations suggest that the hybrid behavior is normal in most of these stars and essentially all $\delta$ Scuti stars are found to show low frequency pulsations (Balona 2010). 

The early studies of pulsating eclipsing binaries found many pulsating Algol (oEA) systems (Mkrtichian 2002, 2003), most of which are $\delta$ Scuti pulsators. Soydugan (2006) made a list of 25 confirmed such systems. Christiansen et al.\ (2007) discovered the first high amplitude $\delta$ Scuti star in an EB. These early studies are nearly all observational, with the detections of a few oscillation frequencies. Thanks to the space missions like {\it CoRoT} and {\it Kepler}, the number of known pulsating EBs increased greatly. The Kepler Eclipsing Binary Catalog has more than 2600 entries, and many of them show signals of pulsations. Recent studies of pulsating EBs  with $\delta$ Scuti/$\gamma$ Doradus components include 
CoRoT 102918586 (Maceroni et al.\ 2013), KIC 11285625 (Debosscher et al.\ 2013), KIC 10661783 (Southworth et al.\ 2011; Lehmann et al.\ 2013), KIC 4844587 (Hambleton et al.\ 2013), DY Aqr (Alfonso-Garzon et al.\ 2014), KIC 3858884 (Maceroni et al.\ 2014), CoRoT 105906206 (Da Silva et al.\ 2014), and KIC 8569819 (Kurtz et al.\ 2015a). These works mostly focus on the binary properties, with mass and radius determination to a few  percent and detections of tens of pulsational frequencies. For a general review of pulsating EBs with other type of pulsators, 
please refer to Huber (2015) and Southworth (2015).

The asteroseismic modeling and mode identification of $\delta$ Scuti stars are notoriously difficult. The fast rotation generally requires 2-D structure models. There are efforts of seismic modeling of single $\delta$ Scuti stars using 1-D stellar models such as Suarez et al.\ (2005) on Altair, where rotation is treated as perturbations to the second order. $\delta$ Scuti stars in eclipsing binaries are rarely modeled. The exception is the work by Maceroni et al.\ (2014) on KIC 3858884, in which the authors identified the g-modes with the Frequency Ratio Method (Moya et al.\ 2005) and identified a possible fundamental radial p-mode with the help of the frequency regularity. The seismic modeling using 2-D models is still the frontier of asteroseismology and mostly adiabatic (Reese et al.\ 2008; Deupree 2011; Ouazzani et al.\ 2012).

KIC 9851944 (K$_p$=11.249, $\alpha_{2000}$=$19$:$56$:$09.732$, $\delta_{2000}$=$+46$:$39$:$40.19$) was first discovered to be an eclipsing binary with an Algol type light curve in the All Sky Automated Survey (Pigulski et al.\ 2009) with an orbital period of 2.1639 days. It was later included in the Kepler Eclipsing Binary Catalog of Pr\v{s}a et al.\ (2011) and Slawson et al.\ (2011) as a detached eclipsing binary. Gies et al.\ (2012) calculated the eclipse times by using the long cadence {\it Kepler} light curve data from quarter $0$ through quarter $9$. They also noted that this system displays near harmonic pulsational variability, possibly on both stars, as evident in their grey scale residual images. Recently, Gies et al.\ (2015) updated their calculations by using all {\it Kepler} quarters. The $O-C$ curve is essentially flat around zero and shows no evidence of a third body or apsidal motion.  They determined the orbital period as 2.16390177 $\pm 0.00000005$ days from the primary eclipses and 2.16390178$\pm 0.00000004$ days from the secondary eclipses. Conroy et al.\ (2014) reported the eclipse times of 1279 {\it Kepler} eclipsing binaries with short periods including KIC 9851944. Armstrong et al.\ (2014) used a model of the binary Spectral Energy Distribution (SED)  to fit the photometry from Everett, Howell \& Kinemuchi (2012), Greiss et al.\ (2012a,b), and 2MASS (Skrutskie et al.\ 2006). They derived the effective temperature of the primary and the secondary as $T_{pri}=6549\pm 409 $K and $T_{sec}=6256 \pm 638 $K.


 
\section{Observations}                        


\subsection{{\it Kepler} Photometry}
 KIC 9851944 was observed by the {\it Kepler} satellite from 2009 to 2013. Short cadence data (sampling rate of $58.8488$ seconds) were obtained during Quarter (Q) $0$, Q$12$ through $14$ and Q$16-17$. In long cadence mode ($29.4244$ minutes sampling), there are data in every quarter from 0 to 17 except for Q7, Q11, and Q15\footnote{The {\it Kepler} Data Release (DR) 23 was used in this paper. Note the short cadence data in the newest DR24 cannot be used in projects requiring high photometric precision. There is another issue related to the smear corrections of the short cadence light curves in all data releases announced on 2016 Feb 5. We evaluated this effect by comparing the target pixel files of the long cadence data (which are not affected) with those of short cadence data and found that this effect is negligible in the case of KIC 9851944. }. 


A preliminary examination of the {\it Kepler} light curves (Figure $1$) shows that this eclipsing binary has evident ellipsoidal variations and pulsations with periods $\approx 2$ hours. The Simple Aperture Photometry (SAP) light curves were used for analysis. The outliers were removed by a 5 sigma criterion. Median differences in quarters were normalized to the same flux. In order to remove the long term trend, a spline curve was used to fit the binned light curve of the out-of-eclipse envelope of the original light curve. This was the method used in Gies et al.\ (2012) and was similar to the de-trending process in Hambleton et al.\ (2013) who used a polynomial fit to the out-of-eclipse envelope of the light curve.  We de-trended the raw light curve of each month and chose the best bin size as the one that minimizes standard deviation of the out-of-eclipse part of the folded light curve. If the dataset contains gaps, we treat each segment separately. We also checked the Fourier transform of the spline trend to make sure we did not remove any intrinsic variations of this system.

\subsection{Ground-based Spectroscopy}
As part of the program of spectroscopic follow up studies to the 41 {\it Kepler} eclipsing binaries in Gies et al.\ (2012), we obtained 10 moderate resolution spectra with the R-C Spectrograph on the Kitt Peak National Observatory (KPNO) 4-meter Mayall telescope and 3 spectra from the DeVeny Spectrograph on the 1.8-meter Perkins telescope at Anderson Mesa of
Lowell Observatory between 2010 and 2011. We used the BL380 grating (1200 grooves mm$^{-1}$) on the R-C Spectrograph at KPNO and this provided wavelength coverage of $3930-4610$\AA. For the DeVeny Spectrograph at Lowell, a 2160 grooves mm$^{-1}$ grating was used and the wavelength range was $4000-4530$\AA.
Both instrument setups provided a resolving power of $R=\lambda/\delta \lambda \sim6000$. The calibration exposures at KPNO used HeNeAr lamps and those at Lowell used HgNeSrCd Pen-Ray lamps. Flat-field and bias spectra were obtained nightly. Standard IRAF\footnote{IRAF is distributed by the National Optical Astronomy Observatory, which is operated by the Association of Universities for Research in Astronomy (AURA), Inc., under cooperative agreement with the National Science Foundation} routines were used to reduce, extract, and calibrate each spectrum. However, wavelength calibrations for spectra from Lowell
Observatory were performed using late-giant stars with known velocities as described in Matson et al.\ (2016) since the available Pen-Ray lamps provide insufficient emission lines for a dispersion solution. Finally, all spectra were normalized to the local continuum and put onto a common heliocentric wavelength grid with uniform $\log \lambda$ spacing. At the observed wavelength range, the mean signal-to-noise ratios (S/N) of the KPNO spectra were about $70-120$ and the Lowell spectra have lower S/N about $30-40$.


The reduced spectra showed that the two components have similar flux contribution and spectral type. As shown in Figure 2, the clear double lines in the composite spectra indicated that the two components are resolved near the Doppler shift extrema.

\section{Data Analysis}  

\subsection{Orbital Elements and Tomographic Reconstruction}         

The observed composite spectra were cross-correlated with two templates for the primary and secondary star to get radial velocities (RVs) and their uncertainties (Zucker 2003) which are listed in Table 1. The templates were constructed from the UVBLUE spectral grid (Rodr\'{i}guez-Merino et al.\ 2005) using preliminary parameters based on the Kepler Eclipsing Binary Catalog of Slawson et al.\ (2011). UVBLUE is a library of theoretical stellar spectra computed with the ATLAS9 and SYNTHE codes developed by R. L. Kurucz, with a spectral resolving power $\lambda/\Delta \lambda= 50,000$. It covers the short wavelength range ($850-4700$\AA) which is ideal for our studies. The RVs derived from KPNO and Lowell spectra have similar statistical uncertainties. However, the Lowell RVs show larger systematic uncertainties which is likely due to the imprecise wavelength calibration.

At first we used templates with atmospheric parameters and projected rotational velocity of $T_{\rm eff},\log g,v \sin i$ =[$6200$K, $4.4$, $23$ km~ s$^{-1}$] for the primary, and
 [$6000$K, $4.4$, 23 km~ s $^{-1}$] for the secondary, with a flux ratio of $F_2/F_1$=$0.72$ and solar metallicity. The templates were later updated to [$7026$K, $4.0$, $52$ km~ s $^{-1}$] and [$6950$K, $3.7$, $70$ km~ s $^{-1}$] for the primary and secondary, respectively, and with a flux ratio of $1.3$ after we performed the preliminary analysis of the separated component spectra (see below).

We derived the orbital elements by fitting the radial velocities with the 
non-linear, least squares fitting program of Morbey \& Brosterhus (1974). The orbital period was fixed to the value in Gies et al.\ 2012 as $P= 2.16390189 \pm 0.00000008$ days, which was the same within uncertainties as from later eclipse time measurements (Gies et al.\ 2015). The period derived from later light curve modeling was almost identical to this value. The adjusted parameters included $T_0$ (time of maximum velocity), $K_1$ (semi-amplitude velocity of the primary), $\gamma$ (system velocity) for the primary star and $K_2$, $T_0$ and $\gamma$ for the secondary. The eccentricity was fixed to $0.0$. We search for a circular orbital solution since the {\it Kepler} light curve suggests that the eccentricity is essentially zero. The system is probably old enough to have completed the circularization. 

The best fitting orbital solutions based on all RVs are shown in Figure $3$. These solutions were used to determine the pixel shifts for each spectrum. We then applied the Doppler tomography program (Bagnuolo et al.\ 1994)  to get the individual spectrum of each component (Fig. 4). In the reconstruction process, we treated the mean flux ratio of the two components in the observed spectral range as a free parameter ($F_2/F_1$), and the reconstructed spectra were subsequently compared with synthetic spectra from the UVBLUE library. The best mean flux ratio was determined as the one that gives the minimum chi square from the spectra residuals of both the primary and the secondary, which was $F_2/F_1=1.34 \pm 0.03$. The spectra from UVBLUE were normalized and convolved with the instrumental broadening and rotational kernel (Gray 2008) before comparing with observational spectra.  

Our goal is to determine the stellar properties of each star from the reconstructed spectra. It is helpful to set the gravity $\log g$ in the analysis from the masses and radii determined from a combined spectroscopic and light curve fit (see section 3.2).

An initial light curve fitting was performed with ELC (see next section for details) and the primary velocity semi-amplitude $K_1$, systemic velocity $\gamma$, and mass ratio were fixed to the values from the radial velocity curve fitting mentioned above. $T_{\rm eff}$ of the primary was initially fixed to the value from Armstrong et al.\ (2014) and adjusted later. We found that in the light curve fitting process, when we changed the primary temperature to 6200K and 7200K, the best fitted inclination only changed by less than 0.4\%, and the $\log g$ only changes by 0.9\%. Thus, the $\log g$ values from the light curve and radial velocity curve modeling were more accurate and so were adopted in fitting the spectra by atmosphere models. A well known problem in spectroscopic analysis is the parameter degeneracy. While the effective temperature and projected rotational velocity $v \sin i$ are usually well constrained, the parameters $\log g$ and metallicity are more difficult to pinpoint and both correlate with $T_{\rm eff}$. Thus setting $\log g$ from the combined analysis will greatly reduce the parameter degeneracy. This procedure was also adopted by Maceroni et al.\ (2014). 

To get the atmospheric parameters of both components, we adopted two different techniques. First, the genetic algorithm {\it pikaia} (Charbonneau 1995) was used. This optimizer is able to explore a broad parameter space and is good at finding the global minimum. The fitting parameters were $T_{\rm eff}$ and $v \sin i$, while the logarithmic metallicity referenced to the Sun [Fe/H] was fixed to $0.0$. 
The parameters were allowed to vary in broad ranges. The  $T_{\rm eff}$ boundaries for the primary and the secondary star were $6500-7500$K and $6200-7200$K, respectively. The $v \sin i$ of both stars were allowed to very from $10$ to $100$ km s$^{-1}$.
In Figure $5$ we show the $\chi^2$ as a function of atmospheric parameters ($T_{\rm eff}$, $v \sin i$) of both stars. The $\chi^2$ has been scaled so that $\chi_{min}^2 \approx \nu$, where $\nu$ is the degree of freedom. We use the intersections of the lower envelope of $\chi^2$ samples and  the level of $\chi_{min}^2+1$ as the $\pm 1\sigma$ parameter bounds, the derived uncertainties are shown in Table $2$. Note this approach probably underestimates the uncertainties due to parameter correlations as the residuals are not independent Gaussian distributions. The $1\sigma$ errors of effective temperatures are quite small ($25$K and $15$K for the primary and secondary, respectively), and we conservatively adopt one tenth of the UVBLUE grid step size ($\Delta T_{\rm eff}=50$K) as the final uncertainties. A comparison of the reconstructed component spectra with synthetic spectra is shown in Figure $4$.

In the second approach, we determine the atmospheric parameters successively (except for the $\log g$, which is always fixed to the value determined from the combined fit). First, a preliminary tomographic reconstruction was performed with the effective temperatures of the two components fixed to [$7026$K, $6900$K] and an estimated mean flux ratio of $1.3$. Since the hydrogen lines are dominated by Stark (pressure) broadening and less sensitive to rotation, we selected six different spectral ranges with least blended metallic lines ($4010-4040$, \ $4041-4080$, $4130-4210$, $4220-4250$, $4260-4290$ \AA ) and determined the best $v \sin i$ by comparing the reconstructed component spectra with a grid of synthetic template spectra of different rotational broadening. The optimal values of $v \sin i$ are found to be [$55.8$ km~s$^{-1}$, $70.5$ km~s$^{-1}$], which are not sensitive to the initial fixed value of effective temperature and flux ratio. Then the $v \sin i$ are fixed to the best values and the spectral separations are performed again over a grid of mean flux ratios, and for each flux ratio we determine the [Fe/H] and effective temperature by minimizing the $\chi^2$ with a Levenberg-Marquardt  algorithm implemented in the package MPFIT (provided by Craig B. Markwardt, NASA/GSFC). The error estimates from MPFIT are formal errors based on the covariance matrix which are probably underestimated. The final results are summarized in Table 2.

The flux ratio and atmospheric parameters from the above two techniques agree very well. 
We adopted the projected rotational velocities from the second method which are very close to the expected synchronized values. Please note the ELC synchronous rate are given in Table $3$. According to Zahn's (1977) theory of radiative damping of dynamic tides for early-type close binaries, the orbital circularization timescale $t_{cir}$ follows the relation (Khaliullin \& Khaliullina 2010): $1/t_{cir}=1/t_{cir1}+1/t_{cir2}$, where $t_{cir,i}=10.5(GM_i/R_i^3)^{0.5}q(1+q)^{11/6}E_{2,i}(R_i/a)^{10.5}$ with $i=1,2$. If we adopt the averaged tidal torque constant $E_2$ of a $1.8 M_\odot$ star on the main sequence $\approx 10^{-8.37}$(Claret 2004), the calculated circularization timescale for a binary system like KIC 9851944 is about $6 \times 10^8$ years. The synchronization timescale is an order of magnitude shorter than the circularization timescale and both timescales are shorter than the age of this system (see the isochrone fitting in Section 4). This system is expected to have synchronized rotation. Our derived $v \sin i$ values, especially those from the metal lines using the Levenberg-Marquardt algorithm, are very close to the synchronized value within uncertainties of about $0.8 \sigma$ and $0.3 \sigma$. In the later binary modeling process, we adopte this synchronization assumption. The observed projected rotational velocities are lower than the average $v \sin i$ in single $\delta $ Scuti stars which is around $120$ km~s$^{-1}$. It is plausible that diffusion can take place more easily for $\delta$ Scuti stars within close binaries and this may explain the observed non-solar metallicity in some close binaries. However, for KIC 9851944 the derived metallicity is essentially within the $1 \sigma$ error box of solar.

 
\subsection{Binary Modeling}     
We use the {\it Kepler} short cadence data to perform the light curve modeling. There are 15 months of SC data, one month in Q0, three months each in Q12, Q13, Q14, Q16 and two months in Q17. We fit the light curve of each month separately as this will account for the possible systematic uncertainties from imperfect de-trending and differences in the photometric aperture definition. The Eclipsing Light Curve (ELC) code (Orosz \& Hauschildt 2000) is used to find orbital and astrophysical parameters for KIC 9851944. ELC utilizes the Roche model and NextGen model atmosphere to synthesize the binary light curve with the effects of gravity darkening
	and reflection included. This code\footnote{This proprietary Fortran code is maintained by Jerome Orosz. Detailed documentation is available on request.} has been used to analyze {\it Kepler} eclipsing binaries (Bass et al.~2012; Sandquist et 
	al.~2013; Rawls et al.~2016), 
	heartbeat stars (Welsh et al.~2011), transiting exoplanets (Wittenmyer et al.~2005),
	and circumbinary planets (Orosz et al. 2012), etc.
	The version revised especially for {\it Kepler} data integrates the stellar Spectral Energy Distribution in the {\it Kepler} passband and also incorporates subtle effects such as contamination, relativistic beaming, and finite integration. ELC can fit the radial velocities (RVs) and the light curves (LC) simultaneously, as done in Williams (2009). 

Due to the sharp difference between the quality of {\it Kepler} photometry and our spectroscopic data, we decide to fit the RVs and LC separately. The RVs are fit first as mentioned in the last section and the corresponding parameters $K_1$ (velocity semi-amplitude), $q$ (mass ratio), and $\gamma$ (system radial velocity) were used to fit the LC data in ELC as these parameters have no effect on the light curves (only very weakly on $q$). After fitting the light curve, the output RV curves from ELC are compared with observed RVs to make sure we have a consistent model.

The original de-trended {\it Kepler} light curve is fit first and the residuals still show strong signals of pulsations (Fig. 6). To pre-whiten the pulsational signal in the light curve, we use the SigSpec package (Reegen 2007) to find significant frequencies in the Fourier spectrum of the residuals down to a signal to noise ratio $\approx 4 $ (spectral significance in SigSpec $\approx 5$). We compare the Fourier spectrum of the residuals in consecutive months, and the peaks that appear in both datasets within the frequency resolution were selected. This can prevent us from selecting frequencies that are due to imperfect de-trending and binary light curve modeling. We only choose the frequencies that have amplitudes larger than about $40$ ppm, because there are many lower peaks below this level that may be unrelated to pulsation. The pulsation signal is then represented by a sum of sinusoids of these selected frequencies and is subtracted from the original {\it Kepler} light curves.

This pre-whitened light curve is fit with ELC again. We use both the genetic algorithm based on {\it pikaia} (Charbonneau 1995) and the Monte Carlo Markov Chain (MCMC) (Tegmark et al.\ 2004) to find the global minimum in parameter space. The parameters for the light curve fitting are $i$ (orbital inclination), $f_1$ (filling factor of the primary), $f_2$ (filling factor of the secondary), $T_{0}$ (time of the secondary minimum)
$\footnote{Throughout the paper, the epoch we adopted is the time of primary minimum $T_0$. The only exception is during the light curve fitting process where ELC uses $T_0$ as the time of secondary minimum.}$
, and \ $T_{\rm eff2}/T_{\rm eff1}$ (effective temperature ratio). The Roche lobe filling factor ($f_1,f_2$) is defined as the ratio of the radius of the star toward the inner Lagrangian point (L1) to the distance to L1 from the center of the star, $f=x_{point}/x_{L_1}$. The filling factors determine the stellar radii, and they are the functional counterpart of the surface effective potential ($\Omega_1,\Omega_2$) in W-D program (Wilson \& Devinney 1971) and PHOEBE (Pr\v{s}a \& Zwitter 2005). The orbital period is also adjusted at first and fixed later on since the converged value is almost identical to the values given by Gies et al.\ (2012).  The primary effective temperature is fixed to the value from spectroscopy as $T_{\rm eff1}=7026K$. The Kepler contamination factor $k$ is the percent of contamination light from other stars in the photometric aperture. In ELC, this effect is accounted for by multiplying the median value of the light curve by $k/(1-k)$. For KIC~9851944, this $k$ factor, taken from the $\it Kepler$ Input Catalog, varies from 0.005 to 0.01 in different quarters and has only negligible effect on the light curve modeling. More than $10^5$ models are calculated and the corresponding parameters and their corresponding $\chi^2$ are recorded. The Markov chains generally converge after about a few $10^4$ iterations. The histogram for each fitting parameter and the correlation of each parameter pair is shown in Figure $7$. 
 
We adopt the final fitting parameters as the average value of all MCMC solutions in different quarters, and their standard deviations as the final systematic error bar. Note the statistical error bars from the light curve fitting in each quarter are much smaller than the systematic error bars we adopted from the quarter-to-quarter differences.

Our final result shows that a circular orbit with two synchronized rotating components can fit both the light curve and radial velocity curve very well. 

We define the distortion to the shape of the star as $D=(R_e -R_p) /R_e$, where $R_p$ and $R_e$ are polar radius and the radius pointing to L1, respectively. The secondary star fills the Roche lobe more and $D$ is much larger (8.5\%) compared with that of the primary (2.7\%).

The circular orbit solution is sufficient, as our final residuals of light curve fit still show pulsations at $10^{-3}$ magnitude level. The circular orbit solution will be undistinguishable with a plausible better fit with very small eccentricity. 
We explore the possibility of smaller eccentricity of this system by examining the published eclipse timings in Gies et al.\ (2015) and Conroy et al.\ (2014). For a circular orbit, the phase difference between the secondary and primary eclipse times ($\delta \phi$) is exactly $0.5$, and a deviation from this value can be shown to be equal to  $\frac{1}{\pi}(1+\frac{1}{\sin i})e \cos \omega$ (Kallrath \& Milone 2009) where $i$ is the orbital inclination. This can give us a lower limit on the eccentricity. For KIC 9851944, the median $0.5-\delta \phi$ of all cycles is $0.000057$ and this suggests $e \ge |e \cos \omega| \approx 0.0001$. Strictly speaking, the time difference between the secondary and primary eclipse times is also affected by the light travel time effect, which is primarily a function of semi-velocity amplitude and mass ratio (see eq. 3 in Bass et al. 2012). Since the mass ratio of KIC 9851944 is very close to 1.0, this light travel effect is very small ($0.3$ seconds) and can be neglected. 

To check if the results from spectroscopy and the binary light curve modeling are consistent (Rozyczka et al.\ 2014), we show the radius ratio of two stars in Figure $8$. We obtain a mean flux ratio $F_2/F_1=1.34 \pm 0.03$ from the blue spectra (at $\approx 4275 $\AA ) in the spectral tomography analysis. Thus, we can calculate the observed flux ratio from the projected areas and surface flux ratio per unit area ($f_2/f_1$) assuming the stars are spherical: $F_2/F_1=(f_2/f_1)(R_2/R_1)^2$, where $f_2/f_1=0.92$  is from the Kurucz models using the atmosphere parameters of the two stars. The radius ratio derived this way is $R_2/R_1=1.22 \pm 0.05$, and this is shown as the red solid line and the corresponding $2 \sigma$ credible region is indicated as the gray shaded region. Another way to estimate the radius ratio is directly from the $v \sin i$ measurements since the system is probably synchronized, so that $R_2/R_1=(v \sin i_2)/(v \sin i_1) =71/56 =1.27\pm 0.29$, and this ratio is indicated as the blue dashed line. The radius ratio from the binary modeling (corresponding to the filling factor ratio) is indicated by the red diamond and the contours. The result $R_2/R_1=1.40$ is larger than that from spectroscopy ($R_2/R_1=1.22$ and $1.27$) indicating a discrepancy. For partial eclipsing binaries, there exists a family of comparable solutions to the light curve modeling. These solutions fall in a valley which satisfies $R_1/a+R_2/a= constant$, this is represented as the black dotted line. We tentatively adopt the radii associated with the best fit of the light curve (Table 3) as it is less model dependent, but we discuss below the implications of solutions with a smaller ratio of $R_2/R_1$.%

 
\section{ Evolutionary and Pulsational Properties}                      

\subsection{Comparision with Evolutionary Models}
The accurate stellar parameters that can be derived from eclipsing binaries offer us opportunities to confront our current stellar structure and evolution theories with observations. We adopted a forward modeling approach and computed non-rotating models with the stellar evolution code MESA (Paxton et al.\ 2011, 2013) with different stellar physics. Convection is described by the mixing length theory (B\"{o}hm-Vitense 1958), with the value of mixing length parameter $\alpha_{MLT}$ fixed to $1.8$. Convective core overshoot is described by the exponentially decaying prescription of Herwig (2000). The OPAL opacity tables (Iglesias \& Rogers 1996) and MESA equation-of-state are used. The default solar mixtures of Grevesse \& Sauval (1998) are adopted as they are close to the solar mixture used in the UVBLUE library. Note that the updated solar mixtures in Asplund et al.\ (2009) have a lower metallicity.

We scanned the mass range from $1.7$ to 1.9 $M_\odot$ in steps of 0.01$M_\odot$, which covers the 1 sigma box of both stars. The exponentially decaying overshooting parameter ($f_{ov}$) was varied in the range from 0.0 to 0.02 in steps of 0.005, which corresponds to the traditional step-wise overshoot parameter of $\alpha_{ov} \in [ 0.0,0.2 ]$ expressed in terms of the local pressure scale height $H_p$. The metal mass fraction (Z) was also varied from 0.01 to 0.02 with a step of $0.002$, with the helium mass fraction fixed to Y= $ 0.28$. Note the solar metal mass fraction $Z_\odot$ we adopted in MESA is $0.02$. 

More than 30000 evolution models were computed. Out of these models, we choose a pair of models with the same age and metal mass fraction (Z) which represent the primary and secondary star, respectively.  All coeval models which fall within the $2 \sigma$ error box  of the observed effective temperature and radius of the two stars have been selected. We use a $\chi^2$ like cost function as the criterion to characterize the goodness of fit:

\begin{displaymath}
\chi^2 = \sum_{i=1}^2 ( (\frac{T_{i} - T_{obs,i}}{\sigma_{T_{obs,i}}})^2 + (\frac {R_{i}- R_{obs,i}}{\sigma_{R_{obs,i}}})^2 + (\frac {M_{i} - M_{obs,i}}{\sigma_{M_{obs,i}}})^2)
\end{displaymath}
 
 Figures 9 and 10 show the $\chi^2$ of a grid of fundamental stellar parameters (mass, effective temperature, radius and age) and two stellar physics parameters (metal mass fraction Z and overshooting parameter $f_{ov}$). The best coeval models have a mass of 1.70$M_\odot$, radius of $2.27 R_\odot$ and effective temperature of $7049$K  for the primary, and 1.81$M_\odot$, $3.19R_\odot$, $6906$K for the secondary, with an age of about $1.25$ Gyr. The corresponding evolutionary tracks and the $\delta$ Scuti and $\gamma$ Doradus instability strips are shown in Figure 11. The primary is a hydrogen-burning main sequence star, which locates it in the middle of the MS phase. The secondary is more evolved and has exhausted the central hydrogen. After a short contraction it is now in the expanding hydrogen shell burning post-MS phase. Only models with metal fraction Z of 0.018 or 0.020 can fit the two data points simultaneously, so this suggests that the bulk metallicity of both stars is close to the solar value. The convective overshooting parameter $f_{ov}$ of the primary star is not well constrained (note the broad lower envelope in Fig. $10$) but the model seems to favor a higher value from 0.010 to 0.015. For the secondary star, no overshooting or low overshooting ($f_{ov}$ less than 0.005) can fit the observations well. The best fitting model pair has a mass ratio of 1.06 which is higher than the observed $1.01 \pm 0.03$ at the $1.6\sigma$ level.
However, we do acknowledge that if the radii are more similar to each other (suggested by spectroscopy; Fig. 8), then a mass ratio closer to $1.0$ does fit the evolutionary tracks in Figure $11$.

We also fit the two stars individually, relaxing the constraints of coevality. The single best-fit models for the primary have a mass range of $1.74-1.75$$M_{\odot}$, an age range of $1.1-1.2$ Gyr, an overshooting parameter of $0.00-0.005$, and metallicity of $0.018$ or $0.02$.  For the secondary star, mass is constrained as $\approx 1.84 M_{\odot}$, age as of $1.2-1.4$ Gyr. The overshooting parameter is poorly constrained, but favors a higher value $0.015-0.02$. The metallicity is also poorly constrained. Both stars can be fitted reasonably well with an isochrone of solar-metallicity.

We also compare the observations with two other stellar evolutionary models: the Dartmouth model (Dotter et al.\ 2008) and Yonsei-Yale ($Y^2$) model (Yi et al.\ 2001). In the $\log g$-$T_{\rm eff}$ plane (Fig. $12$), Dartmouth isochrones of $\approx$ $1.4-1.6$ Gyr can fit the observations of the two stars well, while the best fitting Yonsei-Yale isochrones have ages of $1.2-1.4$ Gyr. In both cases, solar metallicity agrees with the observations well. However, the same mass discrepancy exists: the best fitting Dartmouth isochrone intersect the observation box at a mass of $1.65M_\odot$ and $1.80M_\odot$, this gives an even higher mass ratio of $1.09$. This is not surprising since these two evolutionary models have only fixed physics, while in MESA models we can partially alleviate this discrepancy by evoking different  overshooting parameters in the two stars.

\subsection{Interpretation of Pulsations}

There has been significant advancement in the field of asteroseismology. However, most of these achievements focus on solar-like oscillators (Bedding et al.\ 2011; Beck et al.\ 2011; Mosser et al.\ 2012). The A-F type pulsators, mostly $\delta$ Scuti and $\gamma$ Dor stars still require a better theory to explain observations. Even the first step of asteroseismology, that is mode identification, is notoriously difficult due to our lack of knowledge of crucial stellar physics such as mode excitation, nonlinear effects, and the treatment of rotation.

We analyzed the residuals of the binary light curve to investigate the pulsational properties. We found that masking the eclipses generates strong aliases in the Fourier spectrum and thus the whole residual lightcurves were used in the analysis. A standard pre-whitening procedure was performed with the {\it Period} $04$ package (Lenz \& Breger 2005) to all long cadence data as well as short cadence data with the fitting formula $Z+\sum_i A_i \sin (2\pi (\Omega_i t+\Phi_i))$, where $Z, A_i,\Omega_i, \Phi_i$ are the zero-point shift of the residuals, pulsational amplitudes, frequencies and phases, respectively, and time $t= $ BJD $ - \ 2,400,000$. The calculation was performed to the long and short cadence Nyquist frequencies ($24.47$ d$^{-1}$ and $734$ d$^{-1}$, respectively). No peaks were found beyond the frequency $\approx 25$ d$^{-1}$ in the short cadence spectrum. 
The envelope of the pre-whitened amplitude spectrum was adopted as a conservative noise level. We extracted the final frequencies  from the long cadence data as they have a longer timespan and better frequency resolution. These frequencies have signal to noise ratios (S/N) larger than $4.0$ and were reported in Table 4. We estimated the uncertainties of frequencies, amplitudes, and phases following Kallinger et al.\ (2008). We show the Fourier amplitude spectrum with the window function, the noise spectrum after pre-whitening 89 significant peaks and the extracted frequency peaks in the upper, middle and lower panels of Figure $13$, repectively. 
A remarkable feature in extracted frequencies was that many of them are related to the orbital frequency $f_{orb}=0.46213$ d$^{-1}$ in the form of $f_{i}\pm kf_{orb}$ ($k=1,2,3,...$). We list these frequencies and other combination frequencies in the form of $mf_{i} \pm nf_{j}$ (we restricted to $m,n=1$ or $2$) in the second half of Table 4, while the independent frequencies are listed in the first half.

In the low frequency region ($f < 4$ d$^{-1}$),  the peaks seem to cluster around 
$1.3$ d$^{-1}$ and $2.3$ d$^{-1}$.  Almost all $\delta$ Scuti stars observed by {\it Kepler} show low frequency peaks, this star is no exception. The primary star is located inside the $\gamma$ Doradus instability strip and the secondary star is just hotter than the blue edge of this strip, so these low frequency peaks are possibly g-mode pulsations. 

In the frequency region ($ 4 $ d$^{-1} \le f \le 8 $ d$^{-1}$), there is a quintuplet $f_9,f_{15},f_{22},f_{53}$ around $f_4=5.097$ d$^{-1}$: $f_9=f_4+f_{orb}$, $f_{15}=f_4-f_{orb}$, $f_{22}=f_4+2f_{orb}$, $f_{53}=f_4-2f_{orb}$. In the high frequency region ($ 8 $ d$^{-1} \le f \le 24 $ d$^{-1}$), nearly all the strong peaks are within the range $10$ to $15$ d$^{-1}$, with several lower peaks near $20$ d$^{-1}$. These frequencies correspond to p-mode pulsations of $\delta$ Scuti stars. We find splittings to many of these p-modes including $f_1\rightarrow (f_{24},f_{49},f_{64})$, $f_2\rightarrow (f_{39},f_{42},f_{54},f_{55},f_{68})$, $f_{5}\rightarrow (f_{11},f_{20})$, $f_7\rightarrow (f_{66},f_{75},f_{78},f_{80})$, $f_8\rightarrow (f_{19},f_{27})$, $f_{10}\rightarrow (f_{17},f_{51},f_{74})$, $f_{29}\rightarrow (f_{38},f_{45})$ and $f_{31}\rightarrow (f_{36},f_{48})$ (see the second half of Table 4). These splittings are all related $f_{orb}=0.46213$ d$^{-1}$ and are likely the result of amplitude modulation from eclipses. Due to the different cancellation effects, modes of different spherical degree $l$ have different amplitude modulation. It is possible to identify the modes from these amplitude modulations, the so called {\it eclipse mapping} as in Reed et al.\ (2005) and Biro \& Nulsp (2011). KIC 9851944 has a circular orbit, the tidal effect is from equilibrium tide which is confined to the first and second orbital harmonics. It is surprising to find that $f_{31}=8f_{orb}$, $f_{50}=22f_{orb}$, $f_{77}=50f_{orb}$  and $f_{86}=46f_{orb}$ are large multiple integer times of orbital frequency as such high orbital harmonics are usually found in very eccentric systems such as heartbeat stars (Maceroni et al.\ 2009; Welsh et al.\ 2011; Hambleton et al.\ 2012). Note that da Silva  et al.\ (2014) also find a pulsation frequency at 19 times of orbital frequency in the circular eclipsing binary CoRoT 105906206.

There are other combination frequencies like $f_{23}=f_2+f_3$, these can be explained by nonlinear mode coupling as proposed by Weinberg et al.\ (2013). It is possible to extract information on the mode identification from the combination frequencies (Balona 2012). Recent study emphasizes the importance of combination frequencies as they provide a simple interpretation of the complex spectra of many $\gamma$ Dor and SPB stars (Kurtz et al. 2015b).

As a  first step to understand the theoretical pulsational spectrum of $\delta$ Scuti stars, we show the evolution of pulsational frequencies of $l=0, 1, 2$ modes of a $1.8M_{\odot}$ star from Zero Age Main Sequence (ZAMS) to post-MS phases in Figures $14$ and $15$. A similar diagram can be found in Dupret (2002). These modes suffer from less cancellation effect in broadband photometry like {\it Kepler} and thus are most likely to be observed.
The stellar structure models are calculated from MESA with solar metallicity, overshooting parameter $f_{ov}=0.005$ and helium fraction Y=$0.28$. The pulsating frequencies are calculated with GYRE (Townsend \& Teitler 2013) in the non-adiabatic mode. 

In Figure 14, the star evolves upwards from the bottom of the plot at ZAMS (0.017 Gyr). The fundamental radial mode ($p_1$) and 1st overtone radial mode ($p_2$)(diamonds) have frequencies around $21$ d$^{-1}$  and $27$ d$^{-1}$, with a frequency separation of $6$ d$^{-1}$ (this separation is essentially constant as we move to higher frequency). As the star slowly expands, the frequencies of radial modes decrease monotonically, forming the inclined diamond ridges. At about $1.3$ Gyr, hydrogen in the core is exhausted as it reaches the Terminal Age Main Sequence (TAMS). After TAMS, as mixed modes appear, the spectrum becomes extremely dense (Fig. 15).

As the behavior of radial modes are the simplest, we can compare the position of $l=1$ and $l=2$ modes with the radial ones to get some insight into the relative positions of different modes. For most of the time on main sequence (upper panel in Fig. 14), the positions of $l=1$ dipole modes (orange $p_1$ and $p_2$ dots) are very close to radial modes (black $p_1$ and $p_2$ diamonds) at low radial order. The exception to the closeness due to avoided crossings happens at advanced stages only for very short time intervals. As we move to higher frequency, the l = 1 modes (orange $p_3$ dots) move gradually to the middle
of two consecutive l = 0 diamond ridges ($p_3$ and $p_4$). Similarly, in the lower panel of Figure 14, we can often observe two close l = 0 (diamonds) and l = 2 modes (green dots) at a wide range of frequencies.

All the above discussions support the argument that we can often observe regular frequency separations in $\delta$ Scuti stars. The theoretical mode frequencies in Figure $14$ and $15$ are calculated from non-rotating stellar structure models and are very simplified. $\delta$ Scuti stars usually have fast rotation and the rotational splitting and rotation generated modes can greatly alter the spectrum. However, the regular patterns can be preserved even with fast rotation (Reese et al. 2008). Breger et al.\ (2009) proposed a method to search for regularities by observing the frequency difference histogram. Handler et al.\ (1997) and Garc\'{i}a Hern\'{a}ndez et al.\ (2009, hereafter GH09) searched for regularities by performing a Fourier transform of the observed p-mode frequencies, and the latter authors also assumed all frequencies have amplitudes of unity (hereafter, the FT method). Maceroni et al.\ (2014) applied the method of Breger et al.\ (2009) to the pulsational frequencies of the eclipsing binary KIC~3858884 and identified the position of the fundamental radial mode of the secondary component. Recently, Garc\'{i}a Hern\'{a}ndez et al.\ (2015, hereafter GH15) applied the FT method to  $\delta $ Scuti stars in seven systems which have accurately determined masses and radii  (six eclipsing binaries and the angular resolved star Rasalhague), and they found regular frequency patterns in all of them. The regular frequency spacings are found to be related to the large frequency separation. They also confirmed that the large frequency separation follows a linear relation with the logarithm of the mean density as shown in Suarez et al.\ (2014), and this relation seems to be independent of rotational velocity. We applied the FT method to the independent p-mode frequencies of KIC9851944, and although there seems to be a regularity of $2.3$ d$^{-1}$, which is close to the spacing of consecutive radial modes of the secondary star ($2.4$ d$^{-1}$ as seen in Fig. 16), the result is not very conclusive.

As a preliminary attempt to identify pulsation modes,
we chose representative structure models among the best coeval MESA models which fit the observed $R$, $T_{\rm eff}$ and $M$. Since the models favor a higher mass ratio, we choose $1.70M_{\odot}$ and $1.77M_{\odot}$ as the possible lower and upper mass limits of the primary; for the secondary the limits of $1.79$ and $1.86M_{\odot}$ are adopted. We calculate the non-rotating non-adiabatic frequencies for all models within a $1\sigma$ error box of the observed radius. 

The calculated frequencies need to be corrected for the effect of rotation. To the first order, each $l>0$ mode will split in to $2 l+1$ components with $m=-l, \cdots, l$. The frequencies of the split modes follow the relation: $\omega_{lm} = \omega_{0}+(1-C_{nl})m\bar{\Omega}+O({\bar{\Omega}}^2)$,
where $C_{nl}$ is the Ledoux constant (Ledoux 1951) which depends on the eigenfunction of the mode. $\bar{\Omega}$ is the mean rotational frequency for the mode. For KIC 9851944, the $C_{nl}$ are directly computed in GYRE from mode eigenfunctions. The $l=1,2$ modes of the primary have $C_{nl}$ about $ 0.1-0.3$. For the $l=1,2$ modes of the secondary star, the $C_{nl}$ are about $0.4-0.6$ and $0.2$, respectively.

The relative amplitudes of rotational splitting components to the central $m=0$ mode depend on the inclination of the pulsation axis (Gizon \& Solanki 2003). If the pulsation axis is aligned with the orbital and rotation axis, then at an inclination of $75$ degrees, the $l=1, m=0$ mode has a very small amplitude and the $l=1$ modes with $m= \pm 1$ are more likely to be observed. Similarly, the $l=2, m=\pm 2$ modes and $l=2,m=0$ modes are more likely to be observed.

Both stars in KIC 9851944 rotate at an intermediate value, with $v \sin i \approx 60$ km~s$^{-1}$. Even at this rotation rate, the rotational splitting may already start to deviate from the above simple first order equation (Goupil et al.\ 2000; Dziembowski \& Goode\ 1992; Suarez et al.\ 2006). Here we made an order of magnitude estimation of the second order effect by interpolating the coefficients in Table $1$ in Saio (1981) assuming a polytropic model with $n=3$ following P\'{e}rez Hern\'{a}ndez et al.\ (1995). For the pure $l=1$ p-mode in the observed frequency range, this correction is $\approx 0.03$ d$^{-1}$. A similar estimation for the high order p-modes can be made by using the equation $3.381$ in Aerts et al.\ (2010). For the $l=1$ and $l=2$ p-modes in the observed frequency range, we get similar results, changes of $0.02-0.03 $ d$^{-1}$ for the primary star. 
The distortion due to centrifugal force also alters the oscillation frequencies and it is also a second order effect. We neglect this effect in this analysis as well as the similar effect from the tidal distortion of stars. Another effect of rotation is the mode degenerate coupling (Goupil 2000; Zwintz et al.\ 2014), e.g, between $l=0$ and $l=2$ modes if their frequencies are very close. For low radial orders, the effect is smaller than $\approx 1 \mu$ Hz $= 0.086 $ d$^{-1}$  at $v \le 70 $ km s$^{-1}$ (Goupil 2011). We also neglect this effect in the analysis.

We plot the theoretical frequencies of unstable modes of $l=0, 1, 2$ for the above mentioned representative models and the observed frequencies in Figure 16. Theoretical frequencies of the primary star are from models of $M_1=1.70M_{\odot}$ and $M_1=1.77M_{\odot}$.
Similarly, we show frequencies from models of $M_{2}=1.79M_{\odot}$ and $M_{2}=1.86M_{\odot}$ for the secondary star. Radial, dipole and quadrupole modes are indicated by black, green and red dots, respectively. Due to the extreme denseness of the theoretical frequencies, the rotational splittings are not shown for the secondary star. The symbol size has been scaled to be proportional to the expected mode visibility $S_{nl}$ according to the expressions given by Handberg \& Campante (2011).

The primary star is still on the main sequence, which shows a clear and sparse spectrum. The fundamental to the 2nd or 3rd overtone radial modes are predicted to be unstable. The frequencies above the horizontal red lines have taken into account the $1$st order rotational splitting assuming that the mean rotational frequency $\bar{\Omega}$ is equal to the orbital frequency. The secondary star has an instability range from the fundamental to the the 3rd overtone radial mode. The highest two peaks $f_{1}=10.3997$ d$^{-1}$ and $f_{2}=10.1760$ d$^{-1}$ are likely to be $l=1$ or $l=2$ modes of the secondary. Frequency peaks $f_{18}=19.1267$ d$^{-1}$, $f_{29}=19.4278$ d$^{-1}$ and $f_{67}=20.8235$ d$^{-1}$ are located only in the unstable range of the primary and probably stem from the primary. $f_{18}$ and $f_{29}$ fall into possible range of the second overtone radial mode. $f_{12}$, $f_{10}$ and $f_{3}$ can be the fundamental radial mode of the primary or the second overtone radial mode of the secondary. The high peak $f_{4}$ at $5.0971$ d$^{-1}$ does not seem to be explained by our unstable p-mode frequencies, and could be a g-mode. 
We assume the observed frequencies are from $l=0,1,2$, but it is possible that the $l=3$ or even higher order modes can also be observed. The range of unstable frequencies agrees roughly with the observations. The theory predicts many more excited modes than the observations reveal, but some observed modes are not predicted to be excited. We can see that even with the constrained mass, radius and effective temperature, the mode identification is still difficult.
\section{Conclusions}

Thanks to the unprecedented light curves from the {\it Kepler} satellite, we are discovering more eclipsing binaries with pulsating components. Eclipsing binaries and pulsating frequencies are the only two sources where we can get accurate, model independent fundamental stellar parameters such as mass and radius. Pulsating eclipsing binaries with $\delta$ Scuti and $\gamma$ Doradus stars are the key to our improvement of mode excitation mechanisms in these intermediate mass stars, and this advancement can only be made after we have a statistically large sample of such systems. Here we add one more system to the list, KIC 9851944, two F stars in a circular orbit with a period of 2.1639 days. The two components have similar masses and effective temperatures but very different radii. We try to match the observations with models of different stellar physics and parameters. Both stars probably show $\delta$ Scuti type p-mode pulsations as well as low frequencies pulsations. We made an attempt to understand the general pulsational spectrum of $\delta$ Scuti stars within this mass range. The observed pulsations can be explained by the low order p-modes of the primary and the secondary or the g-mode and mixed modes of the secondary. This work is an effort of preliminary seismic modeling of $\delta$ Scuti stars in eclipsing binaries using 1-D stellar models. We note that even with the mass and radius constrained to $3.9\%$ and $1.3\%$, respectively, the mode identification for $\delta$ Scuti stars from single band photometry of {\it Kepler} is still inconclusive. This is general problem for all $\delta$ Scuti stars. Accurate multicolor photometry and high cadence data of line profile variations will help to partially overcome this difficulty. It is also desirable to model the system with 2-D structure models taking into account the rotational and tidal distortion. The real advancements call for better theoretical understanding of the effects of convection, rotation, tidal interactions and non-linearity on pulsations, which are still the frontiers of asteroseismology.


\acknowledgments 
 
We thank the anonymous referees for helpful comments and suggestions which greatly improve the quality of this paper. We thank Jerome A. Orosz for making his ELC code available to us and for his constant support. We thank Bill Paxton, Rich Townsend and others for maintaining and updating MESA/GYRE. We thank Joyce Guzik, Dean Pesnell, J. C. Su\'{a}rez and D. J. Armstrong for useful discussions. We acknowledge the observations taken using the 4-m Mayall telescope at KPNO and Perkins  telescope at Lowell Observatory. 
This work is partly based on data from the {\it Kepler} mission. {\it Kepler} was competitively selected as the tenth Discovery mission. Funding for this mission is provided by NASA's Science Mission Directorate.
The photometric data were obtained from the Mikulski Archive for Space Telescopes (MAST). STScI is operated by the Association of Universities for Research in Astronomy, Inc., under NASA contract NAS5-26555. This study was supported by NASA grants NNX12AC81G, NNX13AC21G, and NNX13AC20G. This material is based upon work supported by the National Science Foundation under Grant No. ~AST-1411654. A. G. H. acknowledges support from Funda\c{c}\~{a}o para a Ci\^{e}ncia
e a Tecnologia (FCT, Portugal) through the fellowship SFRH/
BPD/80619/2011, and
from the EC Project SPACEINN (FP7-SPACE-2012-312844). Institutional support has been provided from the GSU College 
of Arts and Sciences and the Research Program Enhancement 
fund of the Board of Regents of the University System of Georgia, 
administered through the GSU Office of the Vice President 
for Research and Economic Development.

{\it Facilities:} \facility{Kepler, Mayall, Perkins} 
 


\clearpage
 

\begin{deluxetable}{cccccc}
\tabletypesize{\small} 
\tablewidth{0pc} 
\tablenum{1} 
\tablecaption{Radial Velocities\label{tab1}} 
\tablehead{ 
\colhead{Time}          & 
\colhead{Phase}&
\colhead{$V_{r}$(pri)}        & 
\colhead{$V_{r}$(sec)}    &
\colhead{Observation}                          \\  
\colhead{(BJD-2400000)}           & 
\colhead{}                 &
\colhead{(km s$^{-1}$)} & 
\colhead{(km s$^{-1}$)}& 
\colhead{Source}& 
       
} 

\startdata		

55367.6938 & 0.32 & $-113.5$ $\pm$ 3.2 &105.2  $\pm$ 4.7 & KPNO   \\
 55368.6869   & 0.78 & 117.4  $\pm$ 3.0  & $-117.8$  $\pm$    4.7 & KPNO    \\
 55368.7235   & 0.79 &115.9  $\pm$    3.2 & $-109.6$  $\pm$ 4.8 & KPNO  \\
 55368.7705  & 0.82 &108.5   $\pm$   3.2  &  $-105.1$  $\pm$    4.9 & KPNO \\
 55368.8208   &0.84&95.7   $\pm$     3.0 &   $-99.5$  $\pm$   4.3 & KPNO \\
 55369.6816  &0.24& $-128.9$    $\pm$    3.2   & 114.9  $\pm$    4.8 & KPNO \\
 55369.7259  &0.26& $-126.8$    $\pm$   3.4    &118.7  $\pm$    5.5 & KPNO \\
 55369.7869  &0.29& $-120.9$     $\pm$   3.2    &117.2  $\pm$    4.6 & KPNO \\
 55369.8234  &0.30&$-120.3$    $\pm$   3.1   & 111.5  $\pm$    4.3 & KPNO \\
 55434.8181  &0.34& $-106.6$      $\pm$  2.7   &   99.1 $\pm$   4.1 & KPNO \\
 55449.9141  &0.31& $-101.3$      $\pm$  3.1   & 136.4  $\pm$  4.7 & Lowell \\
 55463.8100   &0.74&129.4       $\pm$ 3.2   & $-104.3$   $\pm$ 5.1 & Lowell \\
 55755.9599   &0.75&106.6       $\pm$ 3.2&$-127.9$   $\pm$ 4.9 & Lowell  

\enddata 
\end{deluxetable}

\newpage 
 
\begin{deluxetable}{lcccc} 
\tabletypesize{\small} 
\tablewidth{0pc} 
\tablenum{2} 
\tablecaption{Atmosphere Parameters\label{tab3}} 
\tablehead{ 
\colhead{Parameter}   & 
\colhead{Primary\tablenotemark{b}}      &
\colhead{Secondary\tablenotemark{b}}      &
\colhead{Primary\tablenotemark{c}} & 
\colhead{Secondary\tablenotemark{c}} 
}
\startdata 
$T_{\rm eff}$ (K)               \dotfill & $7026 \pm 50$ & $6950 \pm 50$
                               & $7018 \pm 76$ & $6881 \pm 70$ \\ 
$\log g$ (cgs)  \dotfill & $3.96\tablenotemark{a}$  & $3.69\tablenotemark{a}$  
                               & $3.96\tablenotemark{a}$  & $3.69\tablenotemark{a}$      \\ 
$v \sin i$ (km s$^{-1}$)     \dotfill & $53 \pm 7$        & $59 \pm 3$    
                               & $56\pm 10$        & $71 \pm 10$            \\ 
$[Fe/H]$ \dotfill & $0.0\tablenotemark{a}$        & $ 0.0\tablenotemark{a}$    
                               & $-0.06\pm 0.05$        & $-0.04 \pm 0.05$             \\ 

\enddata 
\tablenotetext{a}{Fixed.}
\tablenotetext{b}{From genetic algorithm.}
\tablenotetext{c}{From Levenberg-Marquardt algorithm.}
\end{deluxetable} 

\clearpage

\begin{deluxetable}{lccc} 
\tabletypesize{\small} 
\tablewidth{0pc} 
\tablenum{3} 
\tablecaption{Model Parameters\label{tab3}} 
\tablehead{ 
\colhead{Parameter}   & 
\colhead{Primary}      &
 \colhead{Secondary}  &
\colhead{System}       
}
\startdata 
Period (days) & & &$2.16390189\tablenotemark{a}\pm 0.0000008$\\
Time of primary minimum (HJD-2400000) & & &$55341.03987\pm 0.00004$   \\
Mass ratio q=$M_2/M_1$               & & &$1.01\pm 0.03$    \\ 
Orbital eccentricity, $e$              & & &$0.0\tablenotemark{a}$     \\
$\gamma $ velocity (km $s^{-1}$)& & &$-1.3\pm 0.7$\\
Orbital inclination (degree), $i$ & & & $74.52\pm 0.02$\\
Semi-major axis ($R_\odot$), $a$ & & & $10.74\pm 0.14$\\
Mass ($M_\odot$)              & $1.76 \pm 0.07$             & $1.79 \pm 0.07$    \\ 
Radius ($R_\odot$)               & $2.27\pm 0.03$       & $3.19 \pm 0.04$    \\

Filling factor, $f$     &$0.432\pm 0.003$    &$0.627\pm 0.001$\\
Gravity brightening, $\beta$     &$0.08\tablenotemark{a}$    &$0.08\tablenotemark{a}$\\

Bolometric albedo & $0.5\tablenotemark{a}$   & $0.5\tablenotemark{a}$    \\

$T_{\rm eff}$ (K)                    & $7026\tablenotemark{a}\pm 100 $ & $6902 \pm 100$\\
$\log g$ (cgs)  & $3.96\pm 0.03$     & $3.69\pm 0.03$     \\ 
Synchronous $v \sin i$ (km s$^{-1}$)   &   $51.4 \pm 0.7$   &   $72.1 \pm 0.9$              \\ 
Velocity semiamplitude $K$ (km~s$^{-1})$  &$121.9\pm 1.3$ & $120.2\pm 1.7$\\ 
rms of $V_r$ residuals (km~s$^{-1})$  &$6.2$ & $9.7$\\ 
\enddata 
\tablenotetext{a}{Fixed.}
\end{deluxetable}

\begin{deluxetable}{lccccccc} 
\tabletypesize{\small} 
\tablewidth{0pc} 
\tablenum{4} 
\tablecaption{Significant oscillation frequencies\label{tab5}} 
\tablehead{ 
\colhead{}   & 
\colhead{Frequency (d$^{-1}$)}      &
\colhead{Amplitude ($10^{-3}$)}      &
\colhead{Phase (rad/$2\pi$)} & 
\colhead{S/N} &
\colhead{Comment} &
}
\startdata 
$f_{       1}$    & $10.399692 \pm 0.000002$ & $0.653 \pm 0.008$ & $0.343\pm 0.006$ & $132.5$ & $$ \\
$f_{       2}$    & $10.176019 \pm 0.000002$ & $0.548 \pm 0.008$ & $0.764\pm 0.007$ & $114.1$ & $$ \\
$f_{       3}$    & $11.890476 \pm 0.000002$ & $0.454 \pm 0.008$ & $0.249\pm 0.008$ & $100.6$ & $$ \\
$f_{       4}$    & $5.097099 \pm 0.000007$ & $0.404 \pm 0.019$ & $0.002\pm 0.022$ & $35.6$ & $$ \\
$f_{       5}$    & $11.018543 \pm 0.000005$ & $0.229 \pm 0.008$ & $0.124\pm 0.016$ & $49.0$ & $$ \\
$f_{       6}$    & $12.814916 \pm 0.000005$ & $0.223 \pm 0.008$ & $0.259\pm 0.016$ & $50.2$ & $$ \\
$f_{       7}$    & $14.315078 \pm 0.000004$ & $0.216 \pm 0.006$ & $0.493\pm 0.013$ & $60.3$ & $$ \\
$f_{       8}$    & $2.23970 \pm 0.00001$ & $0.210 \pm 0.013$ & $0.373\pm 0.029$ & $27.9$ & $$ \\
$f_{      10}$    & $11.52231 \pm 0.00001$ & $0.202 \pm 0.008$ & $0.998\pm 0.018$ & $45.0$ & $$ \\
$f_{      12}$    & $11.41981 \pm 0.00001$ & $0.153 \pm 0.008$ & $0.565\pm 0.023$ & $34.0$ & $$ \\
$f_{      13}$    & $14.44808 \pm 0.00001$ & $0.137 \pm 0.006$ & $0.852\pm 0.020$ & $39.4$ & $$ \\
$f_{      14}$    & $1.29699 \pm 0.00002$ & $0.127 \pm 0.018$ & $0.190\pm 0.064$ & $12.4$ & $$ \\
$f_{      16}$    & $2.31972 \pm 0.00002$ & $0.112 \pm 0.013$ & $0.091\pm 0.052$ & $15.2$ & $$ \\
$f_{      18}$    & $19.12671 \pm 0.00001$ & $0.098 \pm 0.004$ & $0.104\pm 0.021$ & $38.4$ & $$ \\
$f_{      21}$    & $1.26807 \pm 0.00003$ & $0.084 \pm 0.018$ & $0.350\pm 0.099$ & $8.0$ & $$ \\
$f_{      23}$    & $7.22672 \pm 0.00003$ & $0.079 \pm 0.020$ & $0.242\pm 0.118$ & $6.8$ & $$ \\
$f_{      25}$    & $5.09657 \pm 0.00004$ & $0.066 \pm 0.019$ & $0.573\pm 0.137$ & $5.8$ & $$ \\
$f_{      26}$    & $6.93255 \pm 0.00004$ & $0.063 \pm 0.020$ & $0.173\pm 0.150$ & $5.3$ & $$ \\
$f_{      28}$    & $6.59001 \pm 0.00005$ & $0.059 \pm 0.021$ & $0.992\pm 0.167$ & $4.8$ & $$ \\
$f_{      29}$    & $19.42781 \pm 0.00001$ & $0.058 \pm 0.004$ & $0.124\pm 0.035$ & $22.5$ & $$ \\
$f_{      31}$    & $3.69704 \pm 0.00002$ & $0.056 \pm 0.010$ & $0.196\pm 0.082$ & $9.7$ & $8f_{orb}$ \\
$f_{      32}$    & $2.20292 \pm 0.00003$ & $0.053 \pm 0.013$ & $0.573\pm 0.113$ & $7.0$ & $$ \\
$f_{      33}$    & $2.13439 \pm 0.00003$ & $0.052 \pm 0.013$ & $0.151\pm 0.120$ & $6.7$ & $$ \\
$f_{      34}$    & $5.09768 \pm 0.00005$ & $0.051 \pm 0.019$ & $0.259\pm 0.176$ & $4.5$ & $$ \\
$f_{      35}$    & $1.13657 \pm 0.00005$ & $0.051 \pm 0.019$ & $0.712\pm 0.179$ & $4.5$ & $$ \\
$f_{      37}$    & $11.00534 \pm 0.00002$ & $0.047 \pm 0.008$ & $0.270\pm 0.079$ & $10.1$ & $$ \\
$f_{      40}$    & $4.78514 \pm 0.00005$ & $0.044 \pm 0.016$ & $0.139\pm 0.171$ & $4.7$ & $$ \\
$f_{      41}$    & $14.01095 \pm 0.00002$ & $0.043 \pm 0.007$ & $0.760\pm 0.071$ & $11.2$ & $$ \\
$f_{      44}$    & $14.39802 \pm 0.00002$ & $0.043 \pm 0.006$ & $0.690\pm 0.066$ & $12.1$ & $$ \\
$f_{      46}$    & $8.55119 \pm 0.00003$ & $0.040 \pm 0.010$ & $0.875\pm 0.112$ & $7.1$ & $$ \\
$f_{      47}$    & $11.27238 \pm 0.00003$ & $0.040 \pm 0.008$ & $0.638\pm 0.090$ & $8.8$ & $$ \\
$f_{      50}$    & $10.16680 \pm 0.00003$ & $0.033 \pm 0.008$ & $0.981\pm 0.115$ & $7.0$ & $22f_{orb}$ \\
$f_{      52}$    & $14.21085 \pm 0.00003$ & $0.032 \pm 0.006$ & $0.909\pm 0.090$ & $8.8$ & $$ \\
$f_{      56}$    & $17.27820 \pm 0.00002$ & $0.029 \pm 0.005$ & $0.286\pm 0.075$ & $10.7$ & $$ \\
$f_{      58}$    & $11.78608 \pm 0.00004$ & $0.028 \pm 0.008$ & $0.447\pm 0.129$ & $6.2$ & $$ \\
$f_{      59}$    & $14.49315 \pm 0.00003$ & $0.027 \pm 0.006$ & $0.124\pm 0.101$ & $7.9$ & $$ \\
$f_{      60}$    & $11.43813 \pm 0.00004$ & $0.027 \pm 0.008$ & $0.212\pm 0.134$ & $5.9$ & $$ \\
$f_{      61}$    & $12.35236 \pm 0.00004$ & $0.027 \pm 0.008$ & $0.847\pm 0.138$ & $5.8$ & $$ \\
$f_{      63}$    & $10.50479 \pm 0.00004$ & $0.026 \pm 0.008$ & $0.536\pm 0.150$ & $5.3$ & $$ \\
$f_{      65}$    & $11.42880 \pm 0.00004$ & $0.025 \pm 0.008$ & $0.108\pm 0.145$ & $5.5$ & $$ \\
$f_{      67}$    & $20.82350 \pm 0.00003$ & $0.023 \pm 0.005$ & $0.910\pm 0.097$ & $8.2$ & $$ \\
$f_{      70}$    & $10.40016 \pm 0.00005$ & $0.023 \pm 0.008$ & $0.109\pm 0.172$ & $4.6$ & $$ \\
$f_{      72}$    & $9.59225 \pm 0.00005$ & $0.022 \pm 0.008$ & $0.934\pm 0.174$ & $4.6$ & $$ \\
$f_{      73}$    & $11.36517 \pm 0.00005$ & $0.021 \pm 0.008$ & $0.461\pm 0.168$ & $4.8$ & $$ \\
$f_{      77}$    & $23.10643 \pm 0.00003$ & $0.019 \pm 0.005$ & $0.636\pm 0.113$ & $7.1$ & $50f_{orb}$ \\
$f_{      79}$    & $13.61003 \pm 0.00005$ & $0.019 \pm 0.007$ & $0.636\pm 0.171$ & $4.7$ & $$ \\
$f_{      81}$    & $14.20153 \pm 0.00005$ & $0.018 \pm 0.006$ & $0.645\pm 0.168$ & $4.8$ & $$ \\
$f_{      82}$    & $14.83255 \pm 0.00005$ & $0.015 \pm 0.006$ & $0.387\pm 0.170$ & $4.7$ & $$ \\
$f_{      84}$    & $14.69435 \pm 0.00005$ & $0.015 \pm 0.006$ & $0.090\pm 0.175$ & $4.5$ & $$ \\
$f_{      86}$    & $21.25792 \pm 0.00004$ & $0.015 \pm 0.005$ & $0.208\pm 0.147$ & $5.4$ & $46f_{orb}$ \\
$f_{      88}$    & $21.52354 \pm 0.00005$ & $0.012 \pm 0.005$ & $0.640\pm 0.183$ & $4.4$ & $$ \\
$f_{      89}$    & $21.24588 \pm 0.00005$ & $0.012 \pm 0.005$ & $0.918\pm 0.183$ & $4.4$ & $$ \\

$ $    & $ $ & $$ & $$ & $$ & $$ \\

$f_{      24}$    & $9.93762 \pm 0.00001$ & $0.073 \pm 0.008$ & $0.004\pm 0.051$ & $15.6$ & $f_1-f_{orb}$ \\
$f_{      49}$    & $9.47544 \pm 0.00003$ & $0.035 \pm 0.008$ & $0.468\pm 0.111$ & $7.2$ & $f_1-2f_{orb}$ \\
$f_{      64}$    & $11.32396 \pm 0.00004$ & $0.026 \pm 0.008$ & $0.087\pm 0.139$ & $5.7$ & $f_1+f_{orb}$ \\

$ $    & $ $ & $$ & $$ & $$ & $$ \\

$f_{      39}$    & $9.71393 \pm 0.00002$ & $0.045 \pm 0.008$ & $0.774\pm 0.084$ & $9.5$ & $f_2-f_{orb}$ \\
$f_{      42}$    & $10.63815 \pm 0.00003$ & $0.043 \pm 0.008$ & $0.692\pm 0.091$ & $8.8$ & $f_2+f_{orb}$ \\
$f_{      54}$    & $9.25177 \pm 0.00004$ & $0.030 \pm 0.008$ & $0.736\pm 0.129$ & $6.2$ & $f_2-2f_{orb}$ \\
$f_{      55}$    & $11.10030 \pm 0.00004$ & $0.030 \pm 0.008$ & $0.172\pm 0.125$ & $6.4$ & $f_2+2f_{orb}$ \\
$f_{      68}$    & $11.56241 \pm 0.00004$ & $0.023 \pm 0.008$ & $0.048\pm 0.153$ & $5.2$ & $f_2+3f_{orb}$ \\
$f_{      83}$    & $22.06649 \pm 0.00004$ & $0.015 \pm 0.005$ & $0.694\pm 0.144$ & $5.6$ & $f_2+f_3$ \\

$ $    & $ $ & $$ & $$ & $$ & $$ \\

$f_{       9}$    & $5.55917 \pm 0.00002$ & $0.205 \pm 0.023$ & $0.987\pm 0.052$ & $15.3$ & $f_4+f_{orb}$ \\
$f_{      15}$    & $4.63503 \pm 0.00002$ & $0.118 \pm 0.015$ & $0.526\pm 0.057$ & $13.9$ & $f_4-f_{orb}$ \\
$f_{      22}$    & $6.02135 \pm 0.00004$ & $0.083 \pm 0.023$ & $0.547\pm 0.129$ & $6.2$ & $f_4+2f_{orb}$ \\
$f_{      53}$    & $4.17286 \pm 0.00005$ & $0.031 \pm 0.012$ & $0.043\pm 0.173$ & $4.6$ & $f_4-2f_{orb}$ \\

$ $    & $ $ & $$ & $$ & $$ & $$ \\

$f_{      11}$    & $11.94281 \pm 0.00001$ & $0.169 \pm 0.008$ & $0.732\pm 0.021$ & $37.3$ & $f_5+2f_{orb}$ \\
$f_{      20}$    & $10.09429 \pm 0.00001$ & $0.093 \pm 0.008$ & $0.443\pm 0.041$ & $19.5$ & $f_5-f_{orb}$ \\

$ $    & $ $ & $$ & $$ & $$ & $$ \\

$f_{      66}$    & $15.23934 \pm 0.00003$ & $0.025 \pm 0.005$ & $0.148\pm 0.099$ & $8.0$ & $f_7+2f_{orb}$ \\
$f_{      75}$    & $13.39085 \pm 0.00005$ & $0.019 \pm 0.007$ & $0.472\pm 0.170$ & $4.7$ & $f_7-2f_{orb}$ \\
$f_{      78}$    & $13.85292 \pm 0.00005$ & $0.019 \pm 0.007$ & $0.462\pm 0.168$ & $4.7$ & $f_7-f_{orb}$ \\
$f_{      80}$    & $14.77724 \pm 0.00004$ & $0.019 \pm 0.006$ & $0.639\pm 0.141$ & $5.7$ & $f_7+f_{orb}$ \\

$ $    & $ $ & $$ & $$ & $$ & $$ \\

$f_{      19}$    & $1.31541 \pm 0.00002$ & $0.095 \pm 0.017$ & $0.551\pm 0.085$ & $9.4$ & $f_8-2f_{orb}$ \\
$f_{      27}$    & $1.77756 \pm 0.00003$ & $0.061 \pm 0.015$ & $0.285\pm 0.112$ & $7.1$ & $f_8-f_{orb}$ \\

$ $    & $ $ & $$ & $$ & $$ & $$ \\

$f_{      17}$    & $10.59808 \pm 0.00001$ & $0.098 \pm 0.008$ & $0.794\pm 0.040$ & $19.9$ & $f_{10}-2f_{orb}$ \\
$f_{      51}$    & $11.06020 \pm 0.00003$ & $0.033 \pm 0.008$ & $0.900\pm 0.113$ & $7.0$ & $f_{10}-f_{orb}$ \\
$f_{      74}$    & $11.98450 \pm 0.00005$ & $0.021 \pm 0.008$ & $0.152\pm 0.170$ & $4.7$ & $f_{10}+f_{orb}$ \\

$ $    & $ $ & $$ & $$ & $$ & $$ \\

$f_{      30}$    & $13.52384 \pm 0.00002$ & $0.058 \pm 0.007$ & $0.278\pm 0.056$ & $14.2$ & $f_{13}-2f_{orb}$ \\
$f_{      43}$    & $1.39544 \pm 0.00005$ & $0.043 \pm 0.017$ & $0.963\pm 0.180$ & $4.4$ & $f_{16}-2f_{orb}$ \\

$ $    & $ $ & $$ & $$ & $$ & $$ \\

$f_{      38}$    & $20.35207 \pm 0.00001$ & $0.045 \pm 0.005$ & $0.444\pm 0.049$ & $16.4$ & $f_{29}+2f_{orb}$ \\
$f_{      45}$    & $21.27632 \pm 0.00002$ & $0.041 \pm 0.005$ & $0.963\pm 0.053$ & $15.0$ & $f_{29}+4f_{orb}$ \\

$ $    & $ $ & $$ & $$ & $$ & $$ \\

$f_{      36}$    & $2.77282 \pm 0.00003$ & $0.051 \pm 0.011$ & $0.450\pm 0.102$ & $7.8$ & $f_{31}-2f_{orb}$ \\
$f_{      48}$    & $4.62129 \pm 0.00005$ & $0.040 \pm 0.014$ & $0.117\pm 0.169$ & $4.7$ & $f_{31}+2f_{orb}$ \\
$f_{      62}$    & $8.78964 \pm 0.00004$ & $0.027 \pm 0.009$ & $0.296\pm 0.155$ & $5.2$ & $f_{31}-f_{45}$ \\

$ $    & $ $ & $$ & $$ & $$ & $$ \\

$f_{      57}$    & $10.08109 \pm 0.00004$ & $0.029 \pm 0.008$ & $0.291\pm 0.131$ & $6.1$ & $f_{37}-2f_{orb}$ \\
$f_{      71}$    & $14.93516 \pm 0.00003$ & $0.023 \pm 0.005$ & $0.626\pm 0.113$ & $7.0$ & $f_{41}+2f_{orb}$ \\
$f_{      85}$    & $15.32227 \pm 0.00005$ & $0.015 \pm 0.005$ & $0.396\pm 0.164$ & $4.9$ & $f_{44}+2f_{orb}$ \\
$f_{      76}$    & $13.28659 \pm 0.00005$ & $0.019 \pm 0.007$ & $0.966\pm 0.175$ & $4.6$ & $f_{52}-2f_{orb}$ \\
$f_{      87}$    & $17.74031 \pm 0.00005$ & $0.013 \pm 0.004$ & $0.613\pm 0.163$ & $4.9$ & $f_{56}+f_{orb}$ \\
$f_{      69}$    & $12.35302 \pm 0.00005$ & $0.023 \pm 0.008$ & $0.872\pm 0.158$ & $5.0$ & $f_{65}+2f_{orb}$ \\

\enddata 
\end{deluxetable} 




\begin{figure}
\begin{center} 
{\includegraphics[height=20cm]{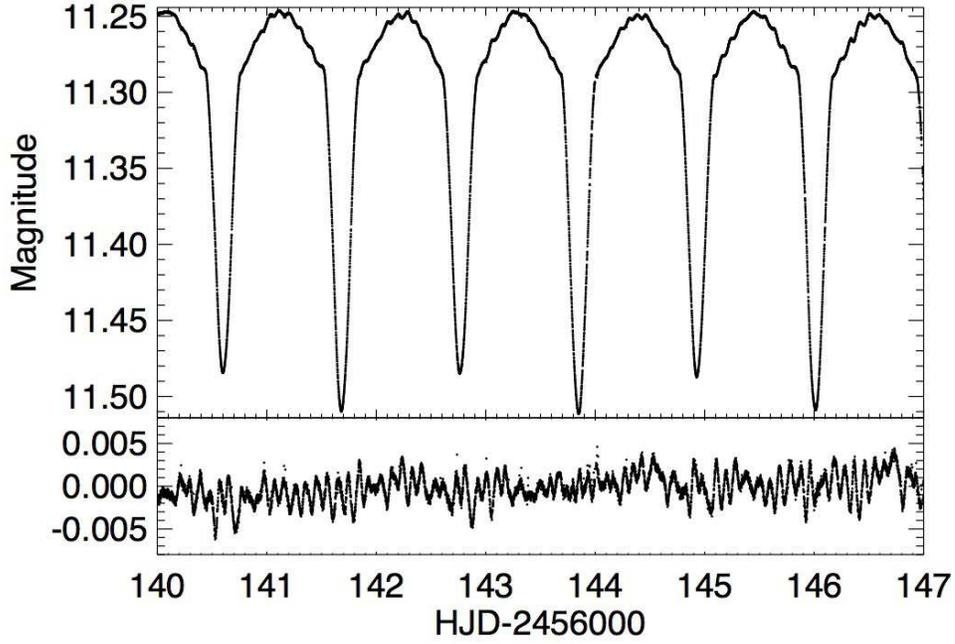}}
\end{center}
\caption{The de-trended light curve of KIC~9851944 during Q13 from short cadence measurements. The lower panel shows the pulsations after subtracting the best binary light curve model. 
\label{fig1}} 
\end{figure} 
 
\begin{figure} 
\begin{center} 
{\includegraphics[height=12cm]{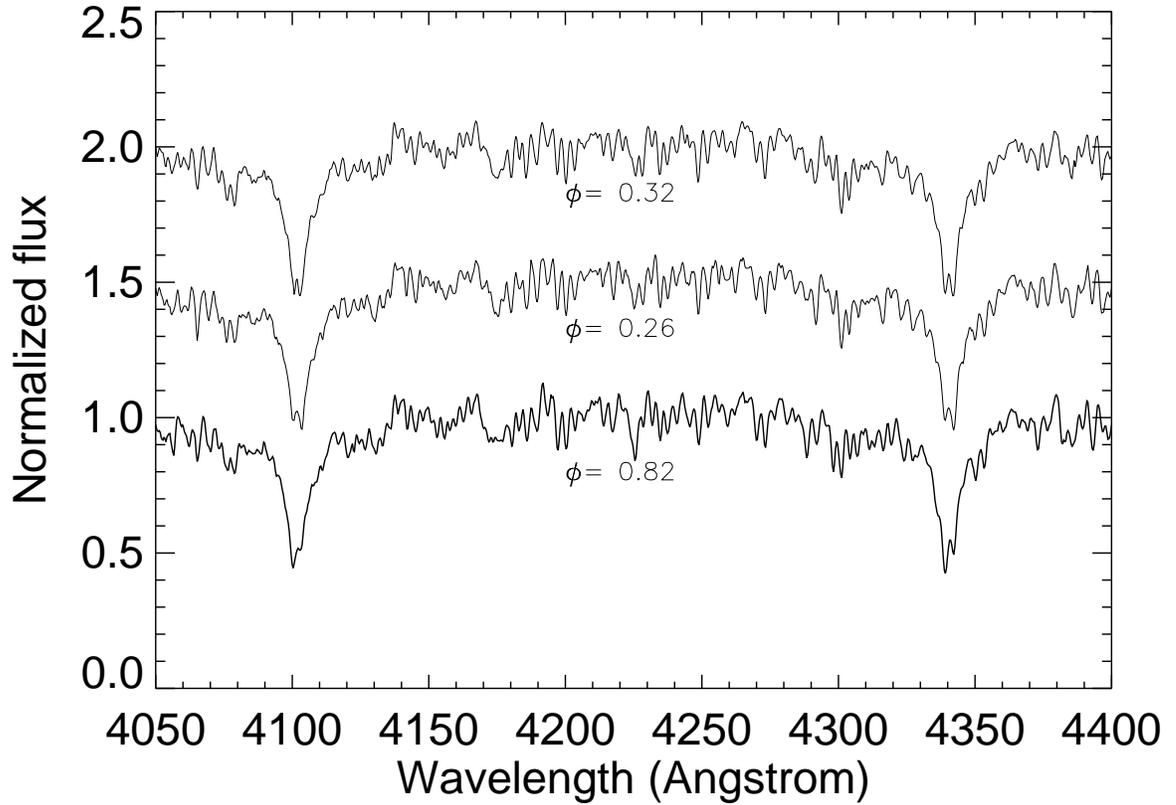}} 
\end{center} 
\caption{The observed composite spectra in the region between H$\delta $ and H$\gamma$ lines. The two components are resolved in the cores of these Balmer lines at times of the velocity extrema. \textbf{The orbital phases ($\phi$) are labeled for each spectrum. For better visibility, the spectra at $\phi=0.26$ and $\phi=0.32$ have been shifted upwards by $0.5$ and $1.0$, repectively.}
\label{fig2}} 
\end{figure} 
 
\begin{figure} 
{\includegraphics[height=12cm]{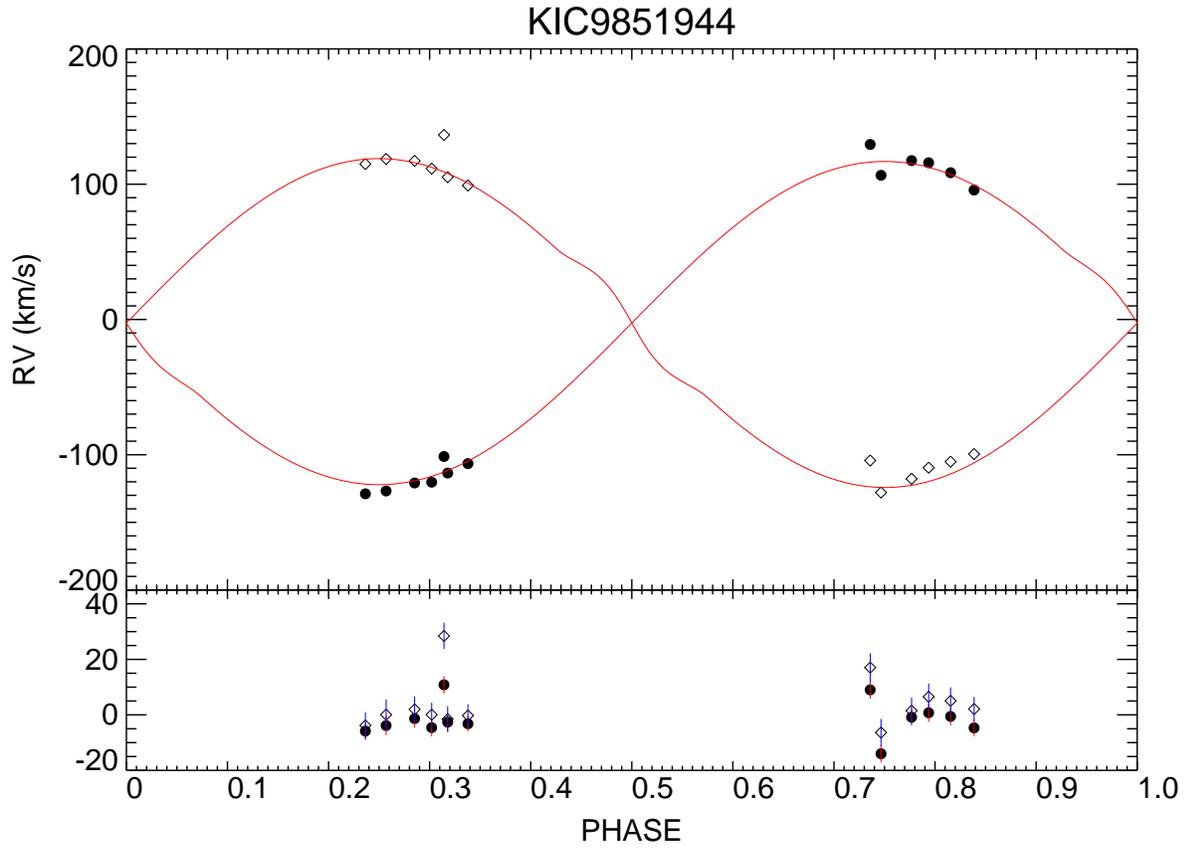}} 
\caption{\textbf{The radial velocities ($V_r$) derived from the cross correlation technique and the best fitting model from ELC.} The primary and secondary are indicated by the filled dots and open diamonds, respectively. The bottom panel shows the residuals.
\label{fig3}} 
\end{figure}

\begin{figure} 
{\includegraphics[height=12cm]{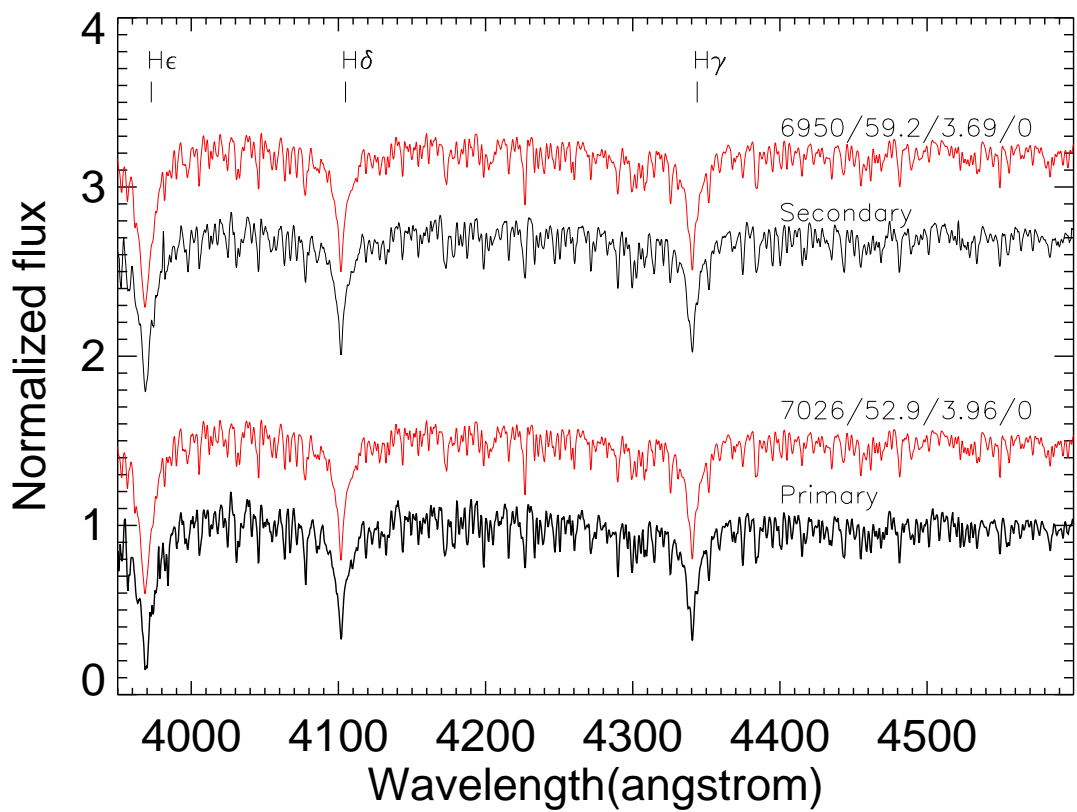}} 
\caption{The reconstructed component spectra of the two components (primary : lower panel; secondary : upper panel) (black) and the corresponding best synthetic spectra from UVBLUE (red). The effective temperature ($T_{\rm eff}$), projected rotational velocity ($v \sin i$), surface gravity ($\log g$) and metallicity ([Fe/H]) are labeled above the synthetic models.
\label{fig4}} 
\end{figure}

\begin{figure} 
{\includegraphics[height=16cm]{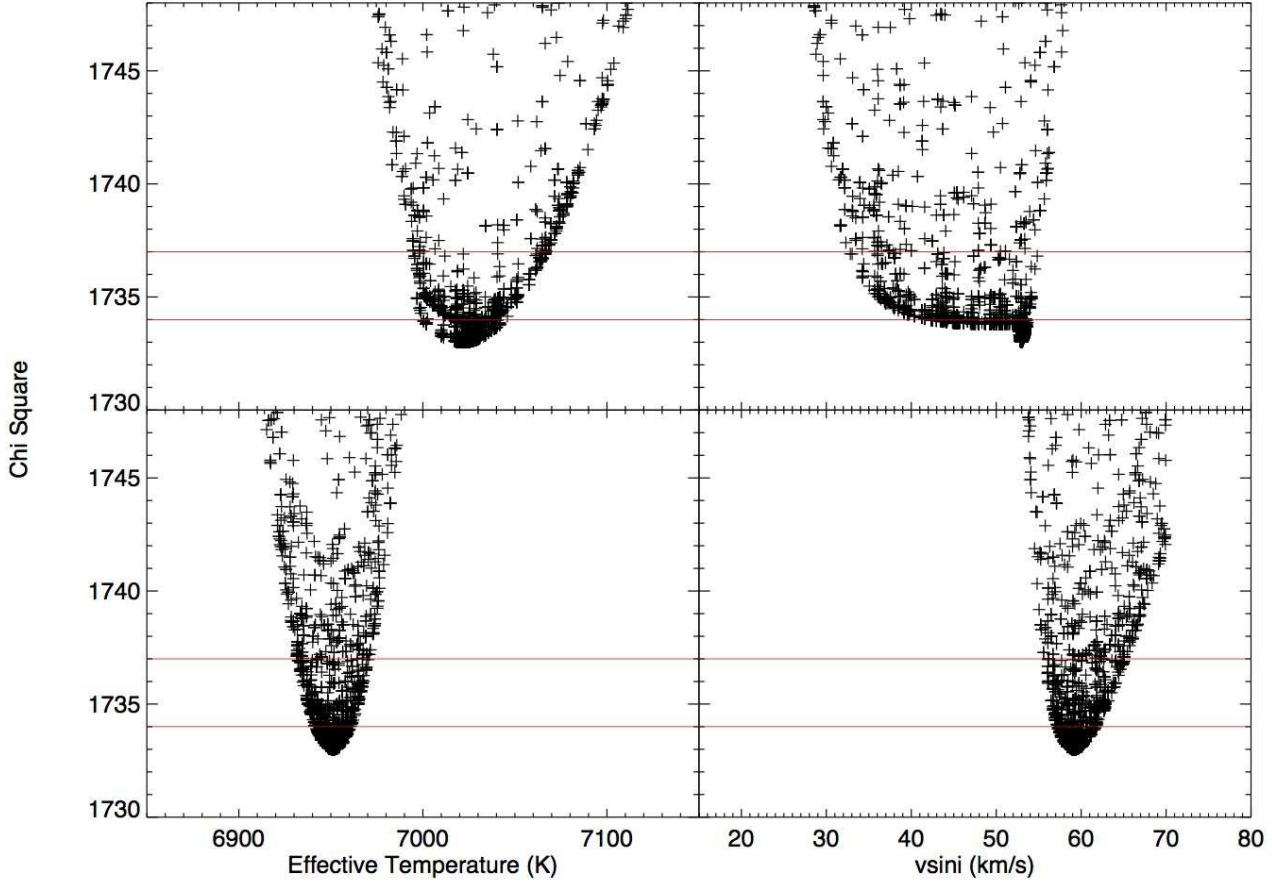}} 
\caption{The $\chi^2$ of stellar atmosphere parameters, $T_{\rm eff}$ and $v \sin i$ from the genetic algorithm. The gravity $\log g$ is fixed to ELC values of $3.96$ and $3.69$, and the metallicity is fixed to the solar value. The $\chi^2$ have been scaled so that $\chi^2_{min} \approx \nu$ \ (the degree of freedom). The two red lines indicate the level of $\chi^2_{min}+1.0$ and $\chi^2_{min}+4.0$. The upper (lower) panels correspond to fits of the reconstructed primary (secondary) spectra.
\label{fig5}} 
\end{figure}

\begin{figure} 
{\includegraphics[height=20cm]{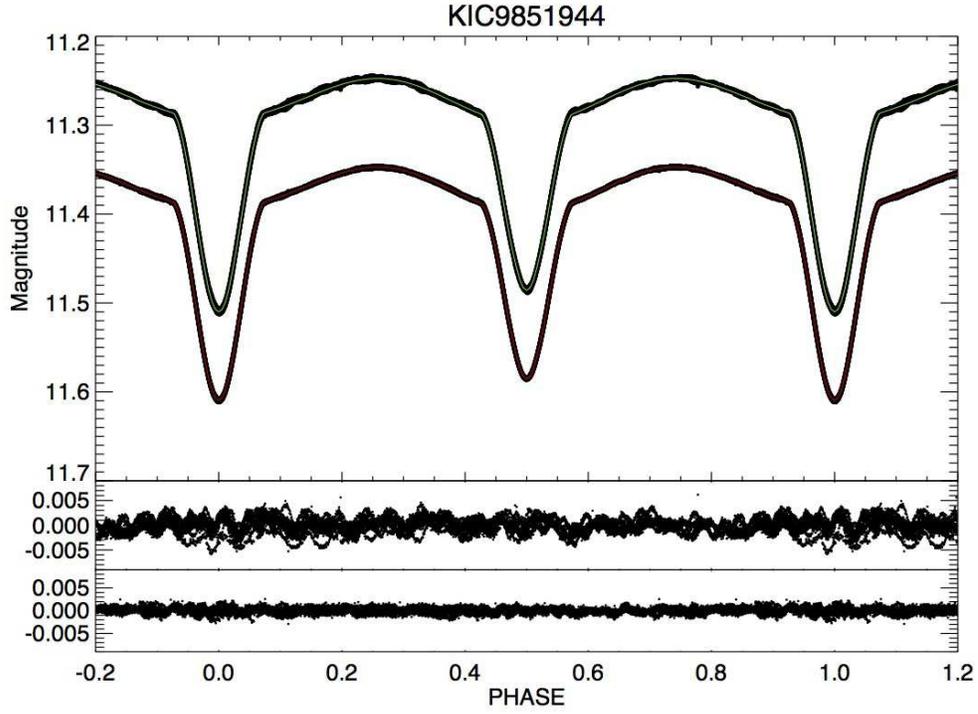}} 
\caption{A fit to the eclipsing binary light curve of KIC9851944 before (above) and after (below) the pre-whitening of pulsations from the dataset of quarter Q$12$b. The model light curves are indicated by the green and red solid lines. The lower two panels show the corresponding residuals.
\label{fig6}} 
\end{figure} 

\begin{figure} 
 {\includegraphics[angle=0,height=16cm]{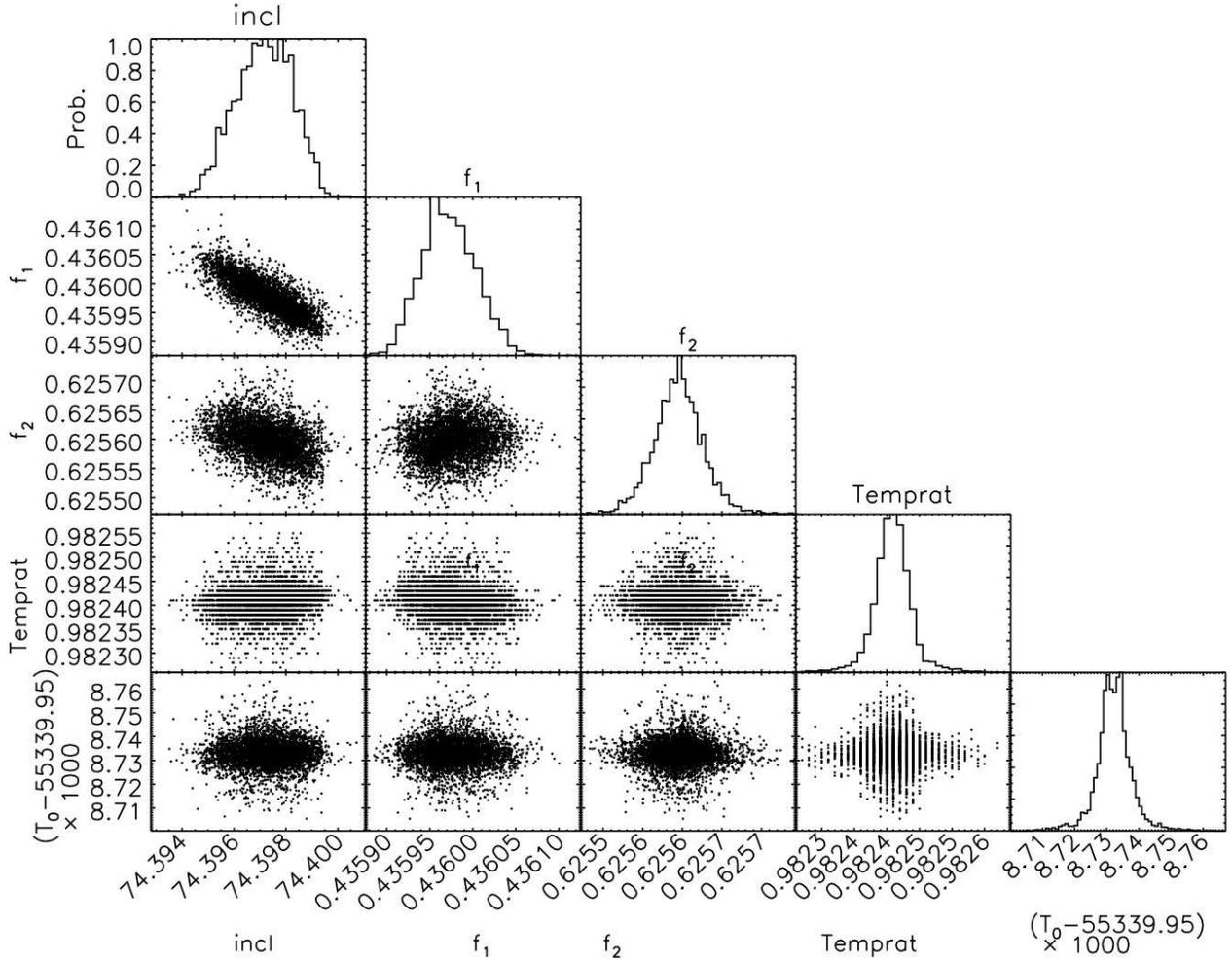}}
\caption{ The parameter correlations from the MCMC analysis of dataset Q$14$. The fitting parameters are $incl$\ (inclination), $f_1$, $f_2$ \ (filling factor), temprat ($T_{\rm eff,2}/T_{\rm eff,1}$) and $T_0$ (time of secondary minimum). The histograms have been normalized to have a maximum peak of unity.
There is a clear correlation between inclination and filling factor, as larger filling factor can be accounted for by a smaller inclination. 
\label{fig7}} 
\end{figure}

\begin{figure} 
 {\includegraphics[height=16cm]{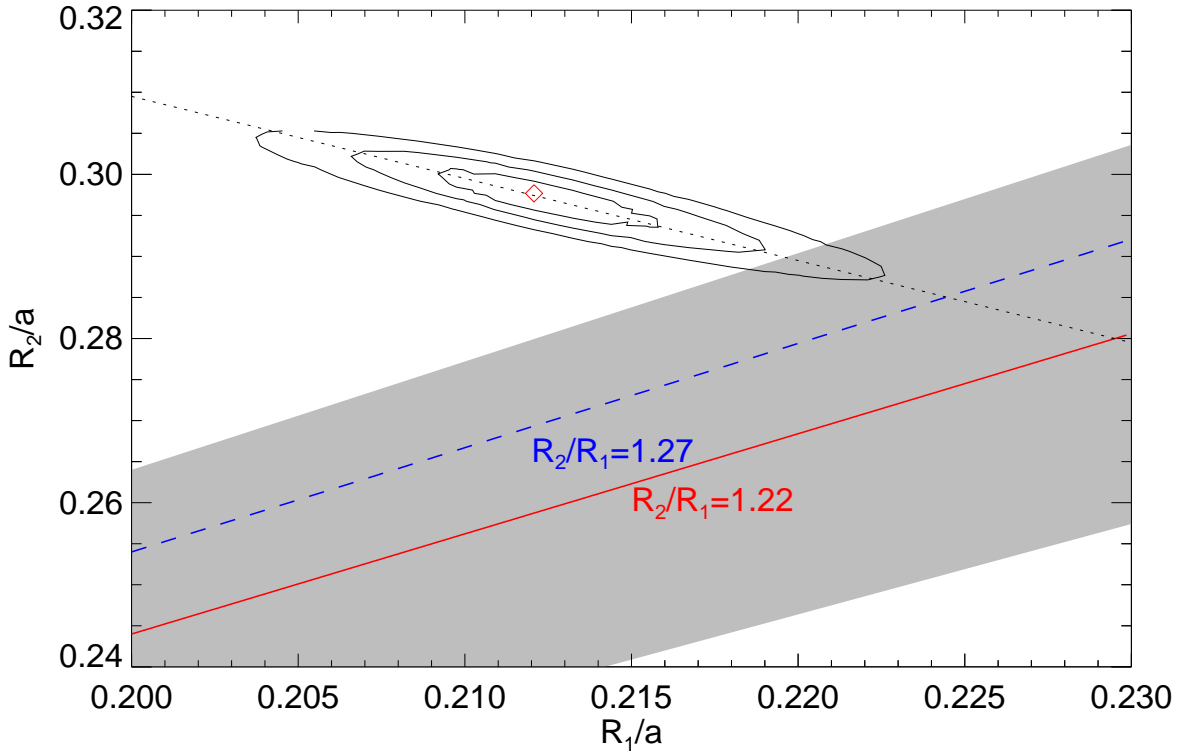}}
 
\caption{ The radius ratio from spectroscopy and binary modeling. The contours show the $1 \sigma$, $2\sigma$, and $3 \sigma$ credible regions of the radius $R_1/a$ and $R_2/a$ from the light curve modeling. The final adopted value is indicated as the diamond. The dark dotted line which crosses the contours corresponds to $R_1/a+R_2/a = 0.51$. It indicates the valley of possible solutions for partial eclipsing systems from the light curve modeling. The radius ratio from spectroscopy is $R_2/R_1=1.22 \pm 0.05$, shown as the red solid line, and the gray shaded area is the corresonding $2\sigma$ credible region. The blue dashed line shows the ratio of $v \sin i$ measurements, $(v \sin i_2)/(v \sin i_1) =71/56 =R_2/R_1=1.27$.
\label{fig8}} 
\end{figure}

\begin{figure} 
{\includegraphics[ angle=90, height=14.3cm]{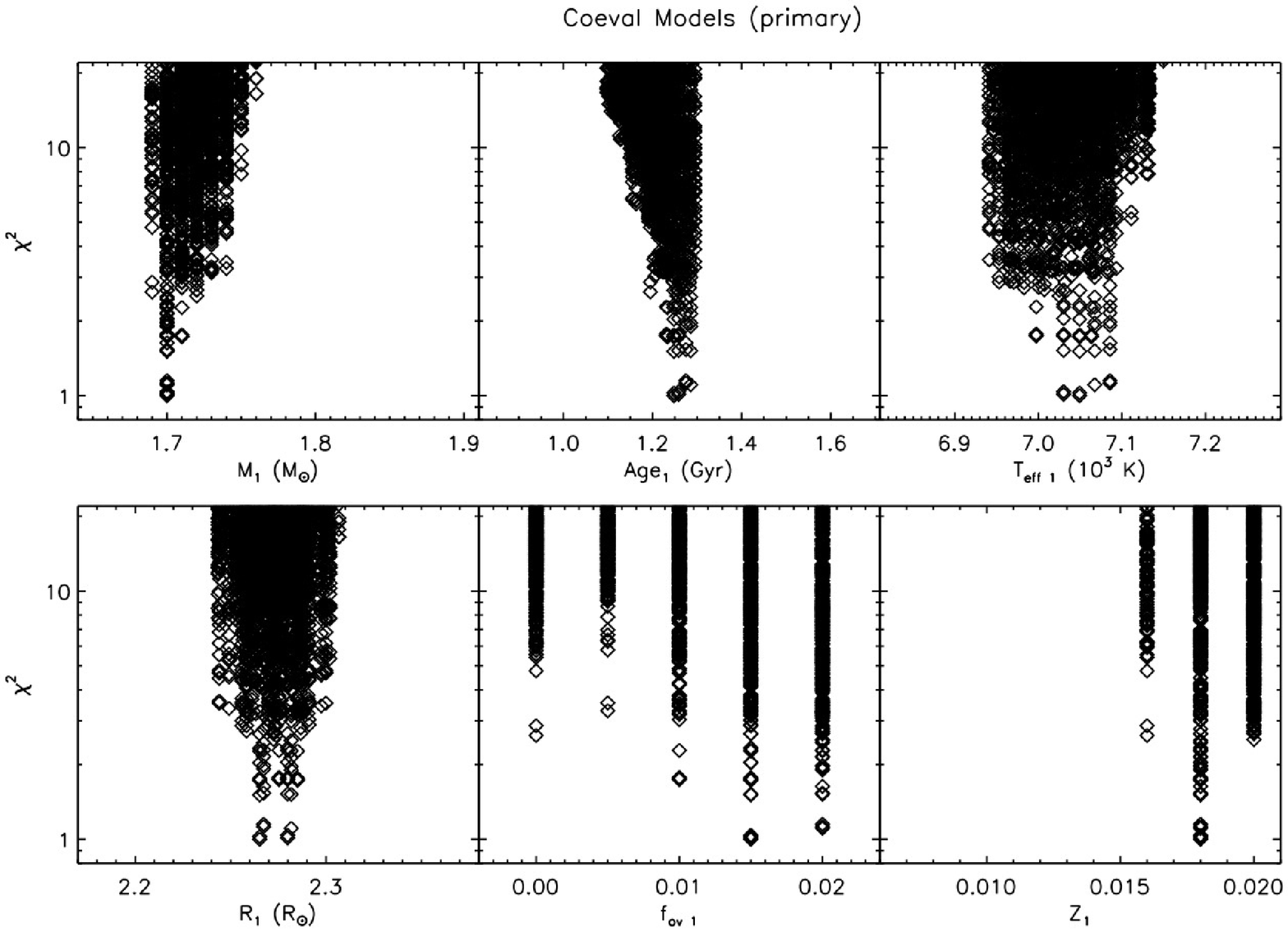}} 
\caption{The distribution of physical parameters of coeval MESA models for the primary star. The minimum of $\chi^2$ has been normalized to $1$.
\label{fig9}} 
\end{figure} 

\begin{figure} 
{\includegraphics[ angle=90, height=14.3cm]{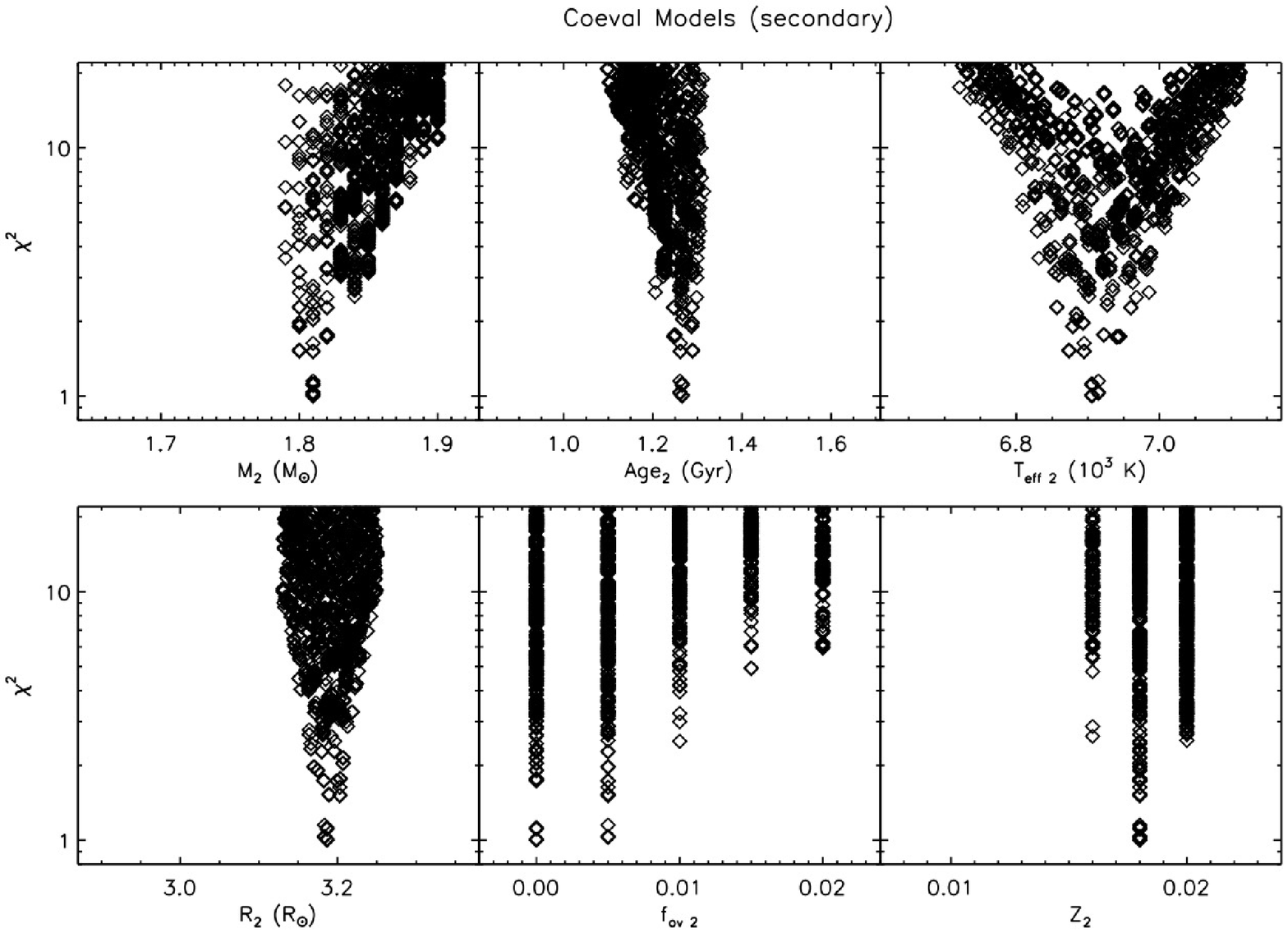}} 
\caption{The distribution of physical parameters of coeval MESA models for the secondary star. The minimum of $\chi^2$ has been normalized to $1$.
\label{fig10}} 
\end{figure}

\begin{figure} 
{\includegraphics[height=12cm]{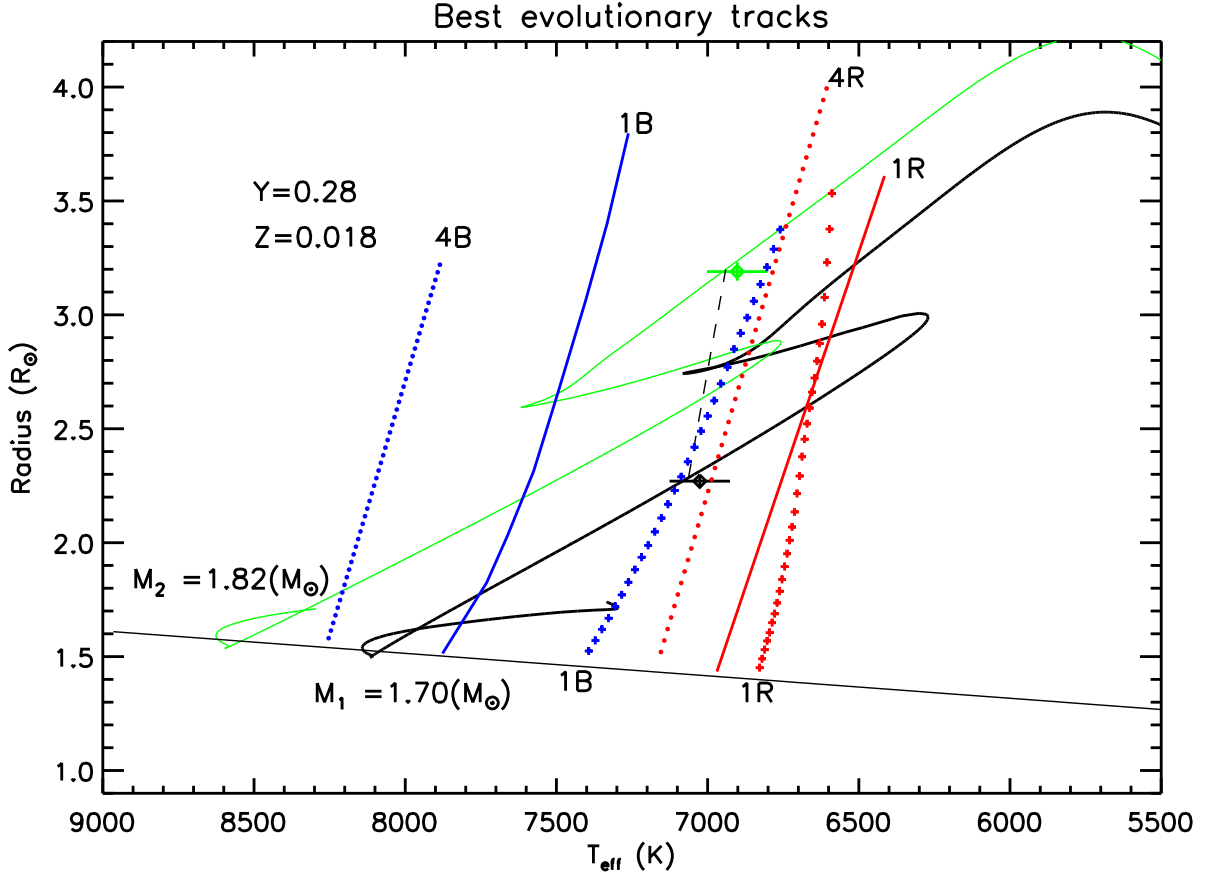}} 
\caption{Best coeval MESA models in the $T_{\rm eff}$-$R$ plane. The evolutionary tracks of the primary ($M=1.70M_{\odot}$) and the secondary ($M=1.82M_{\odot}$) are indicated by the black and green solid curves, respectively. Two diamonds indicate the observational estimates for the primary and secondary stars. The long dash line connecting two models in the tracks represent an isochrone of $1.23$ Gyr. The black solid lines show the Zero- Age Main Sequence. The radial fundamental red and blue edges (1R, 1B) and the $4$th overtone radial red and blue edges (4R, 4B) of $\delta$ Scuti instability strip are indicated by the blue/red solid and dotted lines. The cross lines are the red and blue edges of the $\gamma$ Dor instability strip ($l$ = 1 and mixing length $\alpha_{MLT}$ = 2.0) (Dupret et al. 2005).
\label{fig11}} 
\end{figure}

\begin{figure} 
{\includegraphics[height=12cm]{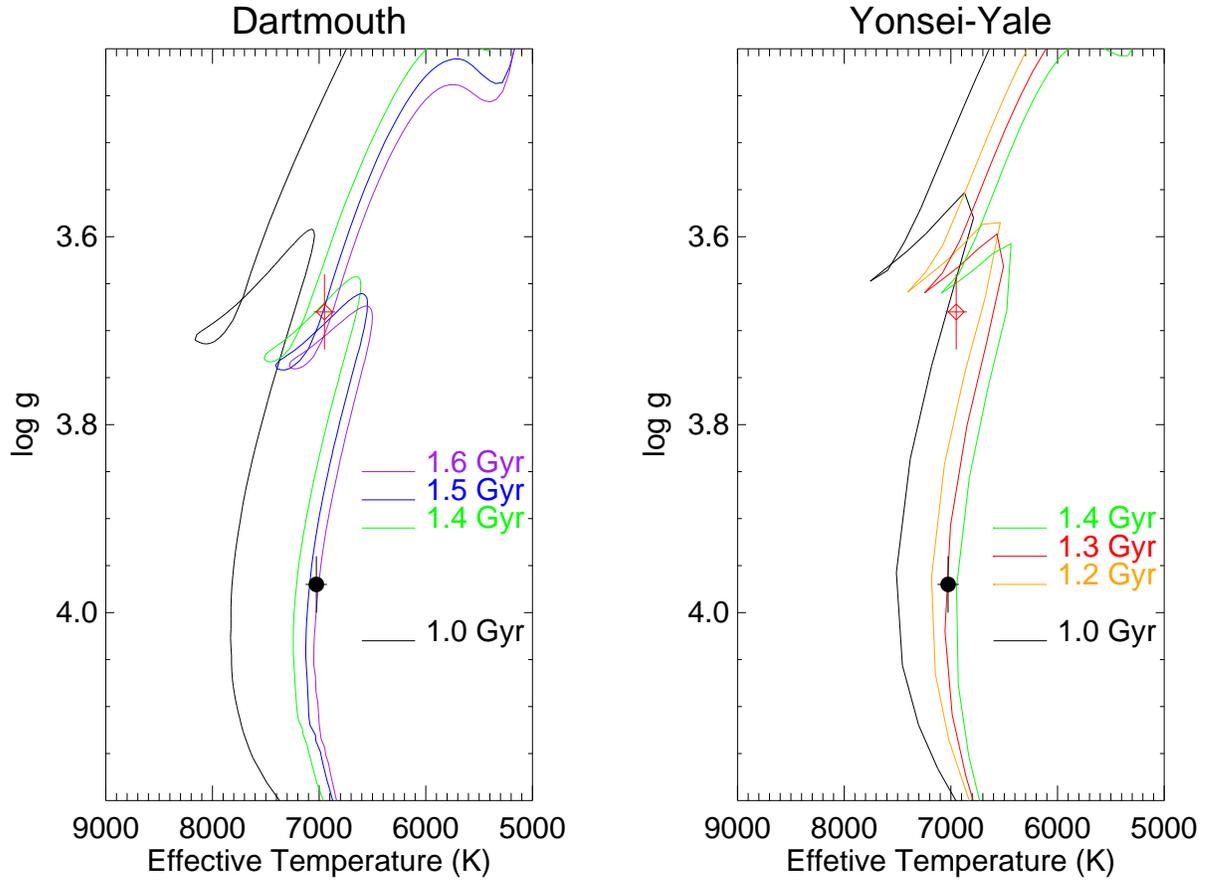}} 
\caption{Comparison of observations with the Dartmouth and Yonsei-Yale isochrones in the $T_{\rm eff}$-$\log g$ plane. The primary and secondary star are indicated by filled dots and open diamonds, respectively.
\label{fig12}} 
\end{figure} 

\begin{figure} 
{\includegraphics[height=12cm]{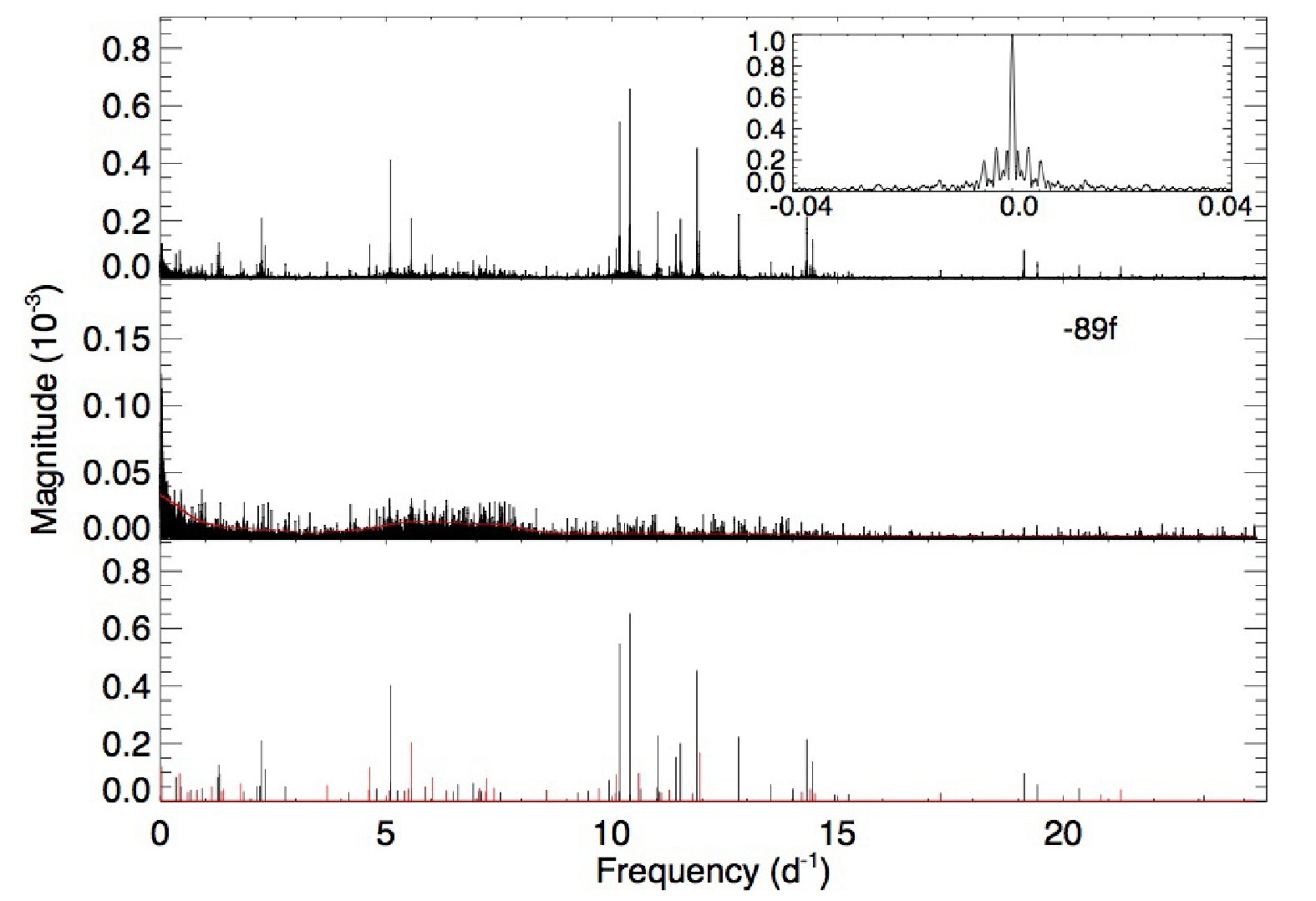}}
\caption{ \textbf{Upper panel:} The amplitude spectrum of the residual light curve of long cadence data (Q$0-10,12,13,14,16,17$) without masking the eclipses. The spectral window is shown in the upper right inset.\textbf{Middle panel:} The spectrum after subtracting $89$ frequencies. The solid red curve represents the adopted noise level. \textbf{Bottom panel:} The extracted significant frequencies with S/N $>$ $4.0$ as listed in Table $4$ (Black: independent frequencies; Red: combination frequencies).
\label{fig13}} 
\end{figure} 

\begin{figure} 
{\includegraphics[height=12cm]{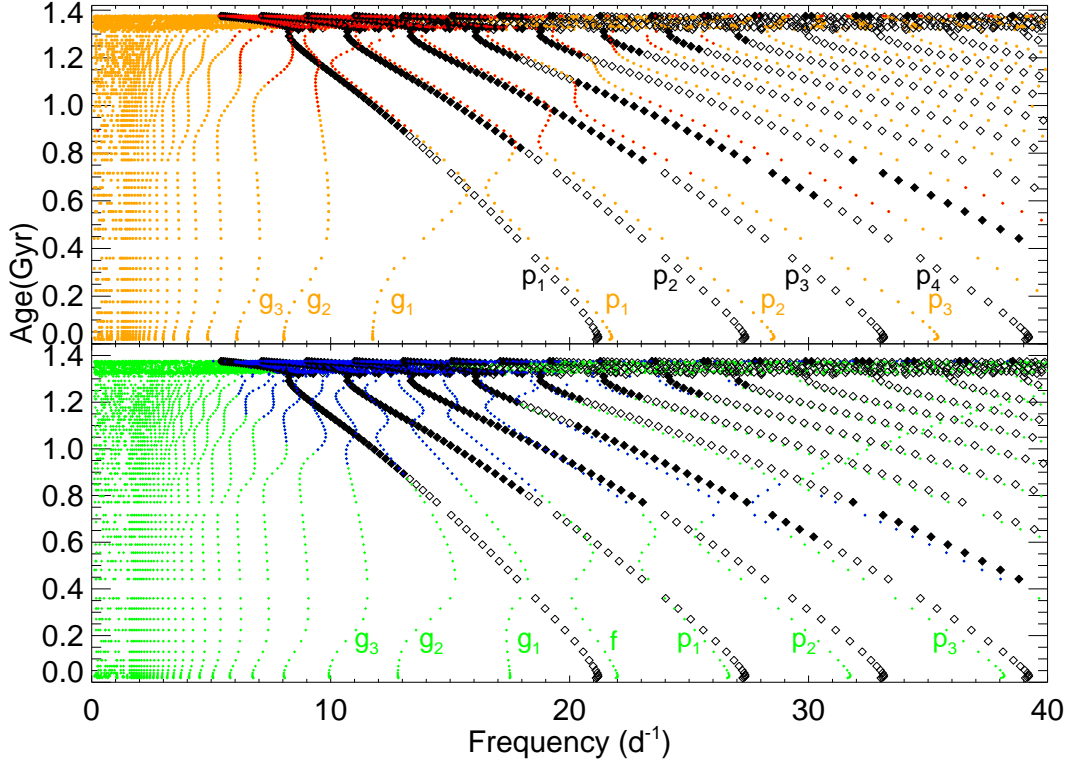}}
\caption{Evolution of oscillation frequencies from ZAMS to post-MS for a $1.8M_{\odot}$ star, with $Z=0.018$, $Y=0.28$, $f_{ov}=0.005$. The upper panel shows the radial  ($l=0$, diamond) and dipole ($l=1$, orange/red dots) modes, the lower panel shows the radial ($l=0$, diamond) and quadruple ($l=2$, green/blue dots) modes. The corresponding radial orders (n) are labeled for p-modes ($p_n$), g-modes ($g_{n}$) and f modes (only for $l=2$). The filled symbols, red dots, and blue dots indicate unstable modes of $l=0, 1 ,2$, respectively. Due to the denseness of high order g-modes, the calculated frequencies less than $\approx 2$ d$^{-1}$ are not reliable.
\label{fig14}} 
\end{figure}

\begin{figure} 
{\includegraphics[height=12cm]{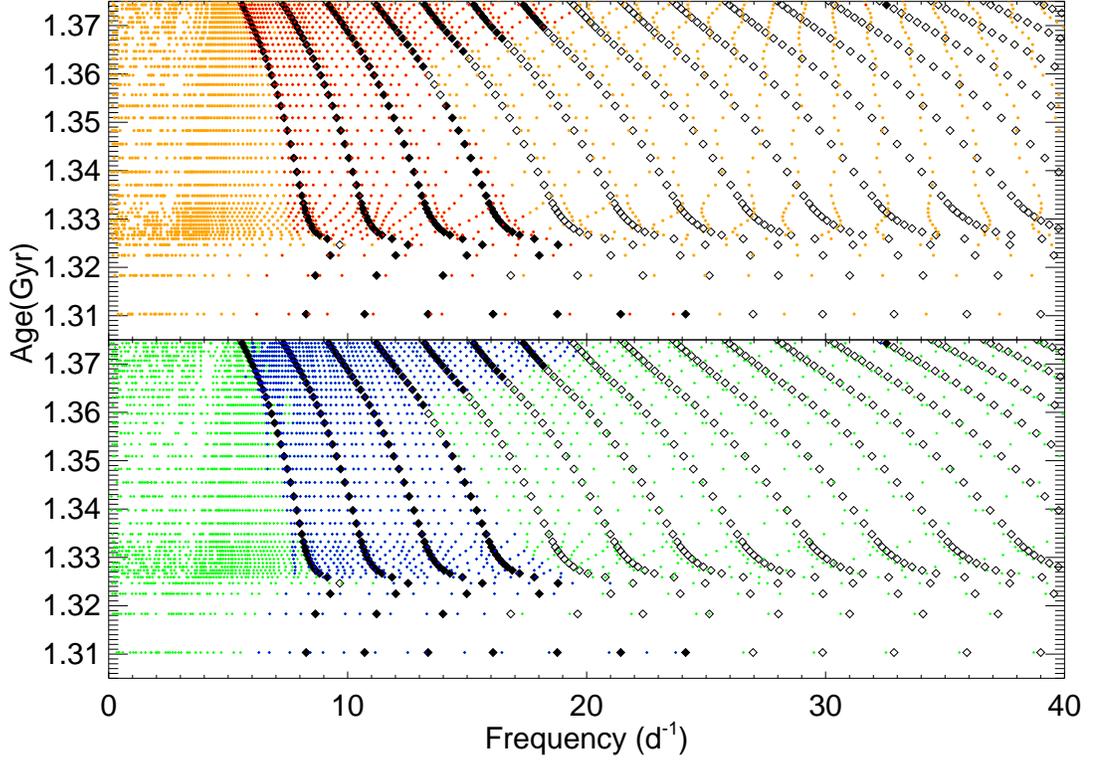}} 
\caption{The evolution of oscillation frequencies from near TAMS to post-MS for a $1.8M_{\odot}$ star. Note the extreme denseness of modes. The upper panel shows the radial ($l=0$, diamond) and dipole ($l=1$, orange/red dots) modes, the lower panel shows the radial($l=0$, diamond) and quadruple ($l=2$, green/blue dots) modes. The filled symbols, red dots and blue dots are unstable modes of $l=0, 1 ,2$, respectively. Due to the denseness of high order g-modes, the calculated frequencies less than $\approx 5$ d$^{-1}$ are not reliable.
\label{fig15}} 
\end{figure}

\begin{figure} 
{\includegraphics[height=12cm]{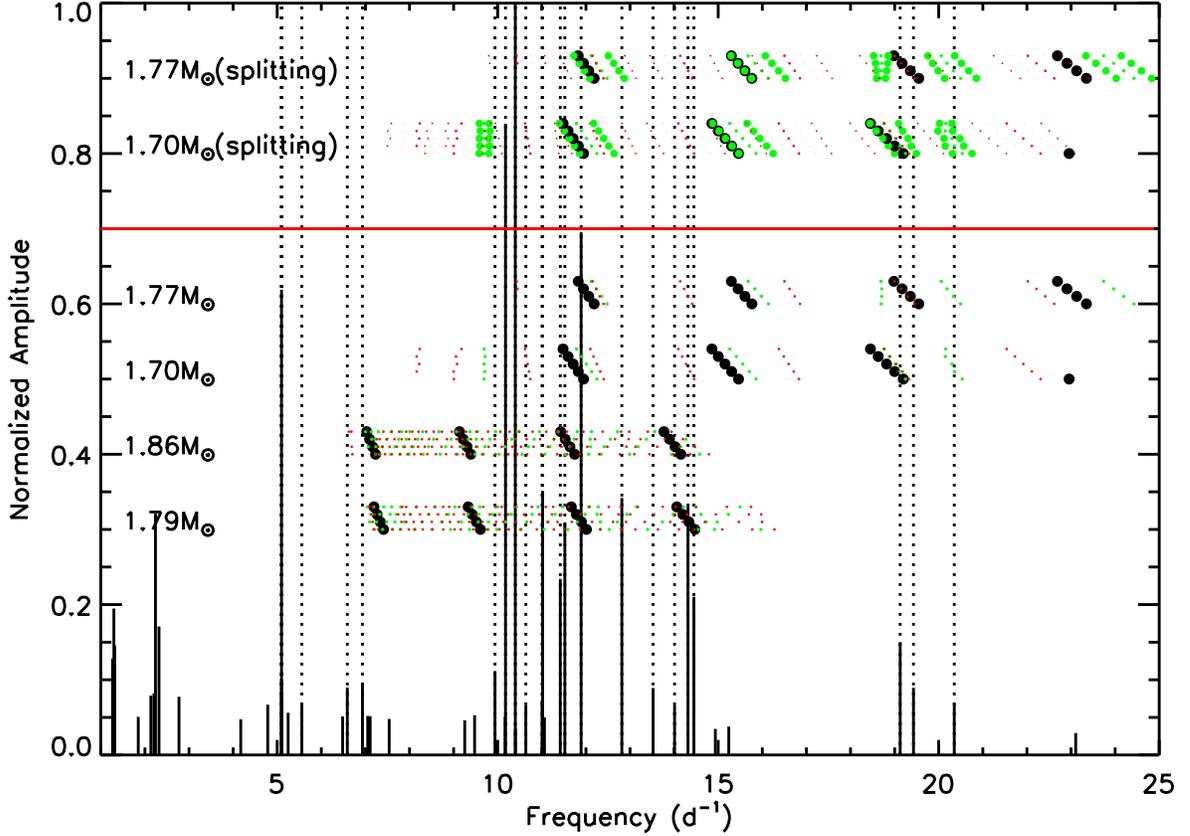}} 
\caption{A comparison of the observed independent frequencies (solid lines, extended as dotted lines for comparison) with theoretical oscillation frequencies (symbols) from models. Theoretical frequencies of the primary star are from models of $1.70$ $M_{\odot}$ and $1.77$ $M_{\odot}$ (lower and upper mass limit) for two cases: $(1)$ the frequencies corrected for the $1$st order rotational splitting (above the horizontal red line); $(2)$ those without rotational splittings (below the red line). The model frequencies of the secondary star are derived from models of $1.79$ $M_{\odot}$ and $1.86$ $M_{\odot}$ (lower and upper mass limit). Note there are four or five models within the $1\sigma$ error box of radius with a fixed mass. Due to the extreme denseness of the modes of the sub-giant secondary, only frequencies without rotational splitting are shown. Black dots are radial modes. Green dots are $l=1$ dipole modes, and $l=2$ modes are indicated as red dots. The symbol size is proportional to theoretical predicted mode visibility (see text).
\label{fig16}} 
\end{figure}

\end{document}